\documentclass[aps,prd,preprint,superscriptaddress,showpacs,ctexart]{revtex4-1} 
\pdfoutput=1 
\usepackage{float}
\usepackage[bottom]{footmisc}
\usepackage{caption}
\usepackage{textcomp}
\usepackage{amsmath}
\usepackage{amssymb}
\usepackage{graphicx}
\usepackage{subfigure}
\usepackage{diagbox}
\usepackage{color}
\usepackage{ulem}
\usepackage{setspace}
\usepackage[unicode=true, bookmarks=false, breaklinks=false,pdfborder={0 0 1},colorlinks=false] {hyperref}

\makeatletter

\newcommand{\bmat}{\left(\begin{array}}
\newcommand{\emat}{\end{array}\right)}
\newcommand{\beq}{\begin{equation}}
\newcommand{\eeq}{\end{equation}}

\makeatletter 
\@addtoreset{equation}{section}
\makeatother  

\def\alt{\mathrel{\mathpalette\gl@align<}}
\def\agt{\mathrel{\mathpalette\gl@align>}}
\def\gl@align#1#2{\lower.6ex\vbox{\baselineskip\z@skip\lineskip\z@
\ialign{$\m@th#1\hfil##\hfil$\crcr#2\crcr\sim\crcr}}}
\def\su5u1{SU(5) \times U(1)}
\def\fsu5u1{SU(5) \times U(1)'}
\def\so10{SO(10)}
\def\sq20{SO(10) \times SO(10)}

\def\bwt{\begin{widetext}}
\def\ewt{\end{widetext}}
\def\be{\begin{equation}}
\def\ee{\end{equation}}
\def\bea{\begin{eqnarray}}
\def\eea{\end{eqnarray}}
\def\bean{\begin{eqnarray*}}
\def\eean{\end{eqnarray*}}
\def\bary{\begin{array}}
\def\eary{\end{array}}
\def\bit{\begin{itemize}}
\def\eit{\end{itemize}}

\makeatother

\begin{document}

\begin{center}

{\Large \bf Monopole operators and symmetry enhancement in ABJM theory revisited   \\}

\end{center}

\vspace{7 mm}

\begin{center}

{ Shan Hu }

\vspace{6mm}
{\small \sl Department of Physics, Faculty of Physics and Electronic Sciences, Hubei University,} \\
\vspace{3mm}
{\small  \sl Wuhan 430062, People’s Republic of China} \\

\vspace{6mm}

{\small \tt hushan@hubu.edu.cn} \\

\end{center}

\vspace{8 mm}

\begin{abstract}\vspace{1cm}

We construct monopole operators for 3d Yang-Mills-matter theories and Chern-Simons-matter theories in canonical formalism. In this framework, monopole operators, although as the disorder operators, could be written in terms of the fundamental fields of the theory, and thus could be treated in the same way as the ordinary operators. We study the properties of the constructed monopole operators. In Chern-Simons-matter theories, monopole operators transform as the local operators with the classical conformal dimension $ 0 $ under the action of the dilation and are also covariantly constant. In supersymmetric Chern-Simons-matter theories like the ABJM model, monopole operators commute with all of the supercharges, and thus are SUSY invariant. ABJM model with level $ k=1,2 $ is expected to have enhanced $SO(8)  $ R-symmetry due to the existence of the conserved extra R-symmetry currents $ j^{AB}_{\mu} $ involving monopoles. With the explicit form of the monopole operators given, we prove the current conservation equation $\partial^{\mu} j^{AB}_{\mu} =0$ using the equations of motion. We also compute the extra $ \mathcal{N}=2  $ supercharges, derive the extra $ \mathcal{N}=2  $ SUSY transformation rules, and verify the closure of the $ \mathcal{N}=8  $ supersymmetry.

\end{abstract}

\maketitle

\begin{spacing}{1.8}
\tableofcontents
\end{spacing}

\section{Introduction and summary}

In any 3d gauge theory with the gauge group containing a $ U(1) $ factor, there is a current $ J^{\mu}=\frac{1}{4\pi}\epsilon^{\mu\nu\lambda} tr F_{\nu\lambda} $ whose conservation is equivalent to the Bianchi identity. The conserved topological charge 
\begin{equation}\label{1q}
Q=\frac{1}{4\pi} \int  d^{2}x \; \epsilon^{ij}tr F_{ij} 
\end{equation}
is called the vortex charge, and the related global symmetry is referred to as the $ U(1)_{J} $ symmetry. Monopole operators are local operators creating (annihilating) the charge $ Q $ \cite{Hoo, Hoo1}. As disorder operators, monopole operators cannot be written as the polynomials of the elementary fields at the insertion point, but dualities can map them onto the operators of that kind \cite{1d,Hoo2, Hoo3,2d,3d,4d,5d}. It is expected that a better understanding of monopole operators can make the 3d dualities more transparent. One such example is the 2d duality relating the massive Thirring model and the sine-Gordon model \cite{kik, kik1}. In the sine-Gordon model, local disorder operators for the creation and annihilation of topological solitons (kinks) are constructed in canonical formalism, while by rewriting the sine-Gordon model in terms of these dual variables, the massive Thirring model is obtained \cite{kik1}. In this paper, we will give a similar construction for monopole operators in 3d, although the main concern is not duality.

Monopole operators also play an important role in membrane models, where the topological charge corresponds to the Kaluza-Klein (KK) momentum along the M-theory circle \cite{tw10,1,2,3,4,5,5a,6,7,7a}. In \cite{tw10}, Aharony, Bergman, Jafferis, and Maldacena (ABJM) proposed an $ \mathcal{N} =6$ $ U(N) \times U(N) $ level $ (k ,-k)$ Chern-Simons matter theory to describe the low energy dynamics of $ N $ coincident $ M2 $-branes probing a $ \mathbb{C}^{4}/\mathbb{Z}_{k} $ singularity. The theory has the manifest $ SU(4) \times U(1)_{J} $ global symmetry, the isometry of $ \mathbb{C}^{4}/\mathbb{Z}_{k}   $, which is expected to be enhanced to $SO(8)  $ when $ k=1,2 $ \cite{tw10}. In AdS/CFT correspondence, ABJM model is dual to M-theory on $AdS_{4} \times S^{7} /\mathbb{Z}_{k}   $ background. With $ S^{7} /\mathbb{Z}_{k}$ viewed as a Hopf fibration over $ \mathbb{CP}^{3} $, supergravity modes carrying KK momenta along the fiber circle should be mapped onto the BPS operators involving monopoles \cite{tw10, tw10a}. In particular, when $ k=1,2 $, twelve dimension-$ 2 $ currents can be constructed from monopole operators, which are conserved by virtue of the dimensions and would enhance the global symmetry to $ SO(8) $ \cite{tw10}. Such expectation is verified in \cite{tw14}. The enhanced $  \mathcal{N} =8 $ supersymmetry is also studied in \cite{tw11, tw12, tw13}.

Monopole operators are usually defined by specifying the singularities of the gauge fields at the insertion point and expanding the quantum fields around this singular background \cite{Hoo1}. For example, in a $ U(N) $ gauge theory, the gauge field singularity is supposed to have the form
\begin{equation}
A_{N/S}=\frac{H}{r}(\pm 1-\cos \theta) d\varphi
\end{equation}
for the north and south charts so that there is some magnetic flux on $ S^{2} $ surrounding it. $ H $ must satisfy the quantization condition $ e^{4\pi i H} =I$ and is an element of the Cartan subalgebra of the form $ H= diag(\frac{m_{1}}{2},\cdots,\frac{m_{N}}{2}) $, $ m_{i}  \in \mathbb{Z}$ \cite{GNO}. For a CFT on $S^{2} \times \mathbb{R}  $, with the specified magnetic flux on $ S^{2} $ given, one can compute quantum numbers like the R-charges and energies of the states, which, by state-operator correspondence, are mapped onto the vortex-charged operators \cite{Hoo1, Hoo2, Hoo3}. In this way, R-charges and conformal dimensions of the vortex-charged operators can be obtained with the quantum corrections taken into account. Since the method only applies for the weakly coupled theories, some strategies must be used to get the strong-coupling result in ABJM theory \cite{tw10a, tw14, S2}.

In this paper, we will return to 't Hooft's original definition of the monopole operator as a singular gauge transformation acting on states \cite{Hoo}. In canonical formalism, such operators can be written in terms of the fundamental variables of the theory, and thus could be treated in the same way as the ordinary operators. This is similar to \cite{kik1} where the kink operator is constructed in 2d sine-Gordon model. In Yang-Mills-matter theories and the Chern-Simons-matter theories, we will give the explicit canonical expressions for the monopole operator $ M_{R} (x)$, which is labeled by the representation $ R $ of the gauge group and is realized as a gauge transformation singular at $ x $. The singular gauge transformation $ M_{R} (x) $, although labeled by $ R $, does not transform in representation $ R $. In Chern-Simons-matter theories, we will define a dressed monopole operator $ \mathcal{M}_{R} (x) $ that would behave as a local operator in representation $ R $ under the local gauge transformation. In non-Abelian gauge theories, $  \mathcal{M}_{R} (x) $ is the monopole operator that is used to build the vortex-charged gauge invariants.

We also study the properties of the constructed monopole operators. We prove the contraction relation (\ref{427f}), which is required in ABJM theory for R-charges to form the $ so(8) $ algebra. We show that in Chern-Simons-matter theories, under the action of the dilation, monopole operators transform as the local operators with the conformal dimension $ 0 $---which, of course, may get quantum corrections from the interactions. We compute the supersymmetry transformation of the monopole operators in ABJM theory. It turns out that monopole operators are invariant under the $ \mathcal{N} =6$ supersymmetry, as well as the $ \mathcal{N} =8$ supersymmetry when $ k=1,2 $. It seems that in supersymmetric Chern-Simons-matter theories, monopole operators always commute with the supercharges, which is not a surprise, since monopole operators are just singular gauge transformations, while the ordinary gauge transformations commute with all gauge invariant operators. On the other hand, in supersymmetric Yang-Mills-matter theories, monopole operators are not SUSY invariant. With the suitable scalar fields added, one may construct BPS scalar-dressed monopole operators preserving part of the supersymmetry.

In ABJM theory, the gauge invariant combination of the monopole operators and the matter fields gives a new set of local operators carrying the vortex charge. In AdS/CFT, part of the KK modes of supergravity on $AdS_{4} \times S^{7} /\mathbb{Z}_{k}$ are mapped onto the vortex-charged operators \cite{tw10}. Since the monopole operators are SUSY invariant and have the classical conformal dimension $ 0 $, these operators could be $ 1/2 $ BPS and thus have the protected conformal dimension matching the spectrum on gravity side.

In the pure Abelian Chern-Simons theory and the $ U(1)_{k} \times  U(1)_{-k} $ ABJM theory, monopole operators can be written as the Wilson lines ending at the inserting point and are covariantly constant \cite{L1, L2, L3, tw10, tw11}. However, when the charged matter is included or the gauge groups are non-Abelian, monopole operators will not be Wilson lines anymore \cite{tw10, tw13}. There is also a controversy on their covariant constancy. In \cite{tw11, tw12}, the covariant constancy condition for monopole operators was assumed when deriving the $  \mathcal{N} =8$ supersymmetry transformation rules in ABJM theory. Whereas in \cite{tw13}, it was shown that such a condition would lead to Eq (6.11), a severe constraint on monopole operators and the scalar fields. In fact, even in \cite{tw11}, for the $ SO(8) $ invariant trial Bagger-Lambert-Gustavsson (BLG) Lagrangian to be identical to the ABJM Lagrangian, the algebraic identities given by Eq (2.7) should also hold, while the closure of the $  \mathcal{N} =8$ supersymmetry requires a few more similar identities. These identities are all constraint equations relating monopole operators and the matter fields.

In ABJM theory, monopole operators commute with the supercharges, while the anticommutator of supercharges gives either a covariant derivative or a field-dependent gauge variation, so from (\ref{824e}), one may conclude that monopole operators are both covariantly constant and invariant under that gauge variation. The latter property could be stated as the algebraic identities for monopole operators and the matter fields. This argument is based on supersymmetry. In section \ref{der}, we will give a proof for the covariant constancy of monopole operators in generic Chern-Simons-matter theories without relying on supersymmetry. We will also show that in Yang-Mills-matter theories, monopole operators are not covariantly constant.

With the explicit form of the monopole operators given, the global symmetry enhancement is revisited. In 3d gauge theories, if the original global symmetry algebra is $ \mathcal{R} $, with the $ U(1)_{J} $ charge $ Q $ added, one may consider the possibility for the enlarged symmetry $\hat{\mathcal{R}} = \mathcal{R}\oplus Q \oplus  \mathcal{R}_{\mathrm{off-diag}} $, where $\mathcal{R}_{\mathrm{off-diag}}  $ denotes off-diagonal elements charged under $ Q $. The form of the global symmetry currents is entirely determined by the field content. The current for $\mathcal{R}_{\mathrm{off-diag}}    $ must involve monopole operators and does not always exist. In the $ k=1,2 $ ABJM theory, $  \mathcal{R}   $ and $\hat{\mathcal{R}}   $ are $ su(4) $ and $ so(8) $ algebras with the currents for $ \mathcal{R}  $ and $  \mathcal{R}_{\mathrm{off-diag}}   $ given by (\ref{jab1}) and (\ref{jab}). We compute the current conservation equation $ \partial^{\mu}j^{AB}_{\mu } =0  $ with the equations of motion plugged in. It turns out that for $\partial^{\mu}j^{AB}_{\mu } =0 $, monopole operators should satisfy the constraints (\ref{89i}). The first is the covariant constancy condition. The rest are the invariance condition under some particular field-dependent gauge variations [stronger than that in (\ref{824e})], which is also the origin of the algebraic identities in \cite{tw11, tw13}. We show that the constructed monopole operators satisfy (\ref{89i}), and so the global symmetry is enhanced to $ SO(8)$.

In fact, in a 3d unitary CFT, the dimension-$ 2 $ currents are always conserved, so to prove the R-symmetry enhancement, it is enough to show $ j^{AB}_{\mu }  $ has the conformal dimension $ 2 $, which is the approach in \cite{tw14}. In our proof of $\partial^{\mu}j^{AB}_{\mu } =0   $, conformal invariance or supersymmetry is not the necessary condition. For example, consider a truncated ABJM model with no fermionic fields, where the monopole operators satisfy the first two equations in (\ref{89i}). Among all $ SU(4) $ invariant scalar potentials, only the quadratic mass term and the sextic potential in ABJM model could make $ \partial^{\mu}j^{AB}_{\mu } =0  $. In this perspective, the symmetry enhancement comes from the classical symmetry of the Lagrangian together with the properties of the monopole operators.

With the extra R-charges added into the $\mathcal{N} =6  $ superconformal algebra of the ABJM theory, two additional supercharges are generated. We get the extra $\mathcal{N} =2  $ supercharges in (\ref{917g}). For the commutation relation (\ref{916g}) to take the presented form, the properties (\ref{89i}) are used again. We write down the extra $\mathcal{N} =2  $ SUSY transformation rules, which are similar to those obtained in \cite{tw11}. We verify the closure of the $\mathcal{N} =8  $ supersymmetry and also discuss the BPS multiplet structure of the theory following the line of \cite{tw15}.

The rest of the paper is organized as follows. In section \ref{mr}, we define the monopole operator $ M_{R}(x) $ as a singular gauge transformation. In section \ref{mr1}, we write down the explicit form of $ M_{R}(x)  $ in Yang-Mills-matter theories and Chern-Simons-matter theories. In section \ref{sec}, we construct the dressed monopole operator $ \mathcal{M}_{R} (x)$ that could transform in representation $ R $. In section \ref{cont}, we derive some contraction relations for $ \mathcal{M}_{R} (x) $. In section \ref{cla}, we compute the classical conformal dimension of $ M_{R}(x) $ and $ \mathcal{M}_{R} (x)$. In section \ref{sus1}, we calculate the supersymmetry transformation of $ M_{R}(x) $ and $ \mathcal{M}_{R} (x) $. In section \ref{der}, we compute the derivative and covariant derivative of $ M_{R}(x) $ and $ \mathcal{M}_{R} (x) $. In section \ref{so}, we prove the current conservation equation $\partial^{\mu}  j^{AB}_{\mu}=0  $ in ABJM theory. In section \ref{SO8}, we study the enhanced $ \mathcal{N}=8 $ supersymmetry. Section \ref{cd} is the conclusion and discussion.

\section{Monopole operator as a singular gauge transformation}\label{mr}

Monopole operators were first introduced by 't Hooft to define an alternative criterion for confinement \cite{Hoo}. The basic idea was to define an operator that creates or annihilates topological charges. It is known that solitons are pure gauge configurations singular at their locations, so the soliton operators can be constructed as the gauge transformations singular at the insertion points. The generic relation between solitons and soliton operators is given in Appendix \ref{AAA1}. Monopole operators in \cite{Hoo} can also have the nontrivial winding $ n $, which excludes the existence of quarks. We will only consider the situation with $ n=0 $ so that the action of the monopole operators on fields in (anti)fundamental representation is also well defined.

Let us start with a brief introduction on group theory. For a group $ G $ with the rank $ r $, $ \{t_{M} | \;  M=1,2,\cdots,\dim G\} $ are generators for the Lie algebra of $ G $ in fundamental representation, among which, $\{ H_{A}|\;A=1,2,\cdots,r\}$ are generators of the Cartan subalgebra. $ tr(t_{M}t_{N})=\frac{1}{2}\delta_{MN} $. Simple coroots are $ r $-dimensional vectors $\{ \vec{\alpha}^{*}_{A} |\;A=1,2,\cdots,r\}$. The irreducible representation $ R $ of the group $ G $ is labeled by $\vec{m}= (m_{1},\cdots, m_{r}) $ with $ m_{A}\in \mathbb{Z} $, $ m_{A} \geq 0 $, $ A=1,2,\cdots,r $, $ m_{1}\geq m_{2} \geq \cdots \geq m_{r} $. When $ G $ and the dual group $ G^{*} $ are identical like $ U(N) $, $ R $ is also in one-to-one correspondence with 
\begin{equation}\label{12}
H_{\vec{m}}=\sum_{A=1}^{r}m_{A}  \vec{\alpha}^{*}_{A}\cdot \vec{H}
\end{equation}
in Cartan subalgebra, where $ \vec{H}=(H_{1},H_{2},\cdots,H_{r})  $. $  \exp \{4 \pi i H_{\vec{m}}\}=I  $. When $G= U(N) $, for an irreducible representation $ R $ labeled by $ \vec{m} =(m_{1},\cdots,m_{N})$, the corresponding $  H_{\vec{m}} $ is 
\begin{equation}\label{24}
H_{\vec{m}}    =diag(\frac{m_{1}}{2},\cdots,\frac{m_{N}}{2}) \;.
\end{equation}

Now consider a 3d gauge theory with the canonical coordinate $( A_{i},\Phi) $ and the conjugate momentum $( \Pi_{i},\Pi) $, $ i=1,2 $. $ A_{i} =A^{M}_{i}  t^{M} $, $\Pi_{i} =\Pi^{M}_{i}  t^{M}  $, $ M=1,2,\cdots, \dim G$. $ A_{i} $ is the gauge field in adjoint representation. $\Phi  $ is the matter field in adjoint or (anti)fundamental representation. The complete orthogonal basis of the Hilbert space can be selected as the eigenstates $ \{\vert A_{i} ,\Phi \rangle | \;\forall \; A, \Phi\} $.

In canonical formalism, it is more appropriate to call the monopole operator the vortex operator since it creates the vortex in 2d space. A vortex at the position $ x $ carrying one unit of the vortex charge $ Q $ can be descried by the gauge configuration $\frac{1}{2}  a_{i}  (x,y)$ with
\begin{equation}\label{16c}
b(x,y)=\epsilon^{ij}\partial^{y}_{i}a_{j}(x,y)=4 \pi  \delta^{2}(x-y)\;,
\end{equation}
where by $ \partial^{y}_{i} $ we mean the derivative with respect to $ y $. $ a_{i}   $ can be solved as 
\begin{equation}\label{24f}
a_{i}(x,y)=2\epsilon_{ji}\frac{x^{j}-y^{j}}{|x-y|^{2}}+\partial_{i}\sigma(y)
\end{equation}
for the arbitrary scalar $ \sigma(y) $. $ a_{i}   $ is not a pure gauge, even though one can still construct some $ e^{i \omega } $ satisfying
\begin{equation}\label{101}
a_{i}(x,y)=-i e^{-i \omega (x,y)   }\partial^{y}_{i} e^{i \omega (x,y)  }\; 
\end{equation}
everywhere except for a singularity at $ x $.

The monopole operator $ M_{R}(x) $ labeled by $ R $ is defined via its action on $ \vert A_{i} ,\Phi \rangle $. When $ \Phi $ is in adjoint representation,
\begin{equation}\label{aqw111}
 M_{R}(x) \vert A_{i}(y), \Phi (y)\rangle   =\vert \Omega_{\vec{m}}(x,y)A_{i}(y)\Omega^{-1}_{\vec{m}}(x,y)- H_{\vec{m}}a_{i}(x,y), \Omega_{\vec{m}}(x,y)\Phi (y)\Omega^{-1}_{\vec{m}}(x,y)\rangle\;.
\end{equation}
When $ \Phi $ is in fundamental representation,  
\begin{equation}\label{121}
 M_{R}(x) \vert A_{i}(y), \Phi (y)\rangle   =\vert \Omega_{\vec{m}}(x,y)A_{i}(y)\Omega^{-1}_{\vec{m}}(x,y)- H_{\vec{m}}a_{i}(x,y), \Omega_{\vec{m}}(x,y)\Phi (y)\rangle\;.
\end{equation}
$ \Omega_{\vec{m}} =  e^{-iH_{\vec{m}}\omega }  $. Moving along a closed curve surrounding $ x $, $ \omega\rightarrow \omega+4\pi $, $  \Omega_{\vec{m}} \rightarrow  e^{- 4\pi i H_{\vec{m}}}  \Omega_{\vec{m}}=\Omega_{\vec{m}} $. $    \Omega_{\vec{m}} $ is single valued and $e^{- 4\pi i H_{\vec{m}}} =I  $ amounts to selecting the winding number $ n=0 $ in \cite{Hoo}. Away from $ x $,
\begin{equation}
   i \Omega_{\vec{m}}^{-1}(x,y) \partial^{y}_{i}\Omega_{\vec{m}}(x,y) =H_{\vec{m}} a_{i} (x,y)
\end{equation}
is satisfied, so $ M_{R}(x) $ is a local gauge transformation everywhere except for a singularity at $ x $.

\section{Monopole operators in canonical formalism}\label{mr1}

It is straightforward to write down the operator expression of $ M_{R} $ in canonical formalism. We will consider two typical situations: $ M_{R} $ in 3d Yang-Mills theory coupling with the matter and $ M_{R} $ in 3d Chern-Simons theory coupling with the matter. Although in these cases, the actions of $ M_{R}   $ on canonical variables are identical, the explicit forms of $ M_{R} $ are different due to the distinct kinetic terms for gauge fields.

In 3d Yang-Mills-matter theory, the Gauss constraint is 
\begin{equation}\label{Omega1}
\Lambda=D_{i}\Pi^{i}-\rho=\partial_{i}\Pi^{i}-i[A_{i},\Pi^{i}]-\rho=0\;,
\end{equation}
where $ \rho $ is the charge density of the matter fields. Local gauge transformation operator $ U(\alpha) $ with the transformation parameter $ \alpha $ is given by 
\begin{equation}\label{22w}
U(\alpha) =\exp \{-i \int d^{2}y \; tr [\alpha(y) \Lambda(y)]\}\;,
\end{equation}
where $\alpha(y)  $ is a Lie-algebra valued function well defined everywhere. As a singular gauge transformation with the parameter $ H_{\vec{m}}\omega $, $ M_{R}(x) $ could be written as 
\begin{eqnarray}\label{1m}
\nonumber M_{R}(x) &=&\exp \{-i \int d^{2}y \; tr [H_{\vec{m}}\omega(x,y) \Lambda(y)]\}\\  &=&
\exp \{i \int d^{2}y \; tr( H_{\vec{m}}[\Pi^{i}(y)a_{i}(x,y)+i [A_{i}(y),\Pi^{i}(y)]\omega(x,y)  +\rho(y)\omega(x,y)   ])\}\;.
\end{eqnarray}

In a 3d Chern-Simons-matter theory with the level $ k $, the canonical commutation relation for the gauge field is \cite{CSS} 
\begin{equation}\label{23c}
[A^{M}_{i}(x),A^{N}_{j}(y)]=\frac{2 \pi i}{k}\delta^{MN}\epsilon_{ij}\delta^{2} (x-y)\;, 
\end{equation}
and the Gauss constraint is 
\begin{equation}\label{Omega2}
\Lambda=\frac{k}{4\pi}\epsilon^{ij}F_{ij}-\rho=\frac{k}{2\pi}\epsilon^{ij}(\partial_{i}A_{j}-\frac{i}{2}[A_{i},A_{j}])-\rho=0\;.
\end{equation}
The monopole operator $ M_{R}(x) $ is given by
\begin{eqnarray}\label{2m}
\nonumber M_{R}(x) &=&\exp \{-i \int d^{2}y \; tr [H_{\vec{m}}\omega(x,y) \Lambda(y)]\}\\ \nonumber &=&
\exp \{\frac{ik}{2\pi} \int d^{2}y \; tr (H_{\vec{m}}[ \epsilon^{ij}A_{j}(y)a_{i}(x,y)+\frac{i}{2}\epsilon^{ij}[A_{i}(y),A_{j}(y)]\omega(x,y) +\frac{2\pi}{k}\rho(y) \omega(x,y) ])\}\;.\\
\end{eqnarray}

In (\ref{24f}), $  a_{i}$ is determined up to a local gauge transformation, $ a_{i} \sim  a_{i}+\partial_{i} \sigma $. For $ M_{R}(x) $ in (\ref{1m}) and (\ref{2m}), with $ \omega $ replaced by $ \omega +\sigma $, we will get $M'_{R}(x)  \sim U(H_{\vec{m}}\sigma) M_{R}(x)    $ with $U(H_{\vec{m}}\sigma)   $ an ordinary local gauge transformation. In particular, in $ U(1) $ pure Chern-Simons theory,  
\begin{equation}
M(x)=\exp \{\frac{ik}{4\pi} \int d^{2}y \;  [ \epsilon^{ij}A_{j}(y)a_{i}(x,y) ]\}\;.
\end{equation}
One can always select the suitable $ \sigma $ so that $ a_{i} $ is nonvanishing only at a Dirac string $y(s) $ with $ 0 \leq s < \infty $, $y(0)= x $, $ y(\infty)= \infty$. In this case, 
\begin{equation}
M(x)=\exp \{ ik \int^{\infty}_{0}ds\;    A_{i}[y(s)]\dot{y}^{i}(s) \}
\end{equation}
is a Wilson line starting from $ x $ and extending to infinity, as expected \cite{L1, L2, L3}.

Similar to the local gauge transformation, the action of $ M_{R} $ on canonical fields is 
\begin{equation}
M_{R}(x)A_{i}(y)M^{-1}_{R}(x)=e^{i H_{\vec{m}}  \omega(x,y)}A_{i}(y)e^{-i H_{\vec{m}} \omega(x,y) }+H_{\vec{m}} a_{i}(x,y)\;,
\end{equation}
\begin{equation}
M_{R}(x)\Pi_{i}(y)M^{-1}_{R}(x)=e^{i H_{\vec{m}}  \omega(x,y)}\Pi_{i}(y)e^{-i H_{\vec{m}} \omega(x,y) }\;,
\end{equation}
and 
\begin{equation}
M_{R}(x)\Phi(y)M^{-1}_{R}(x)=e^{i H_{\vec{m}} \omega(x,y)}\Phi(y)
\end{equation}
or 
\begin{equation}
M_{R}(x)\Phi(y)M^{-1}_{R}(x)=e^{i H_{\vec{m}} \omega(x,y)}\Phi(y)e^{-i H_{\vec{m}} \omega(x,y) }
\end{equation}
for $ \Phi $ in fundamental representation or adjoint representation. However, for the field strength
\begin{equation}
F = \epsilon^{ij}F_{ij}=\epsilon^{ij}(\partial_{i}A_{j}-\partial_{j}A_{i}-i[A_{i},A_{j}])\;,
\end{equation}
we have 
\begin{equation}\label{Q}
M_{R}(x)F(y)M^{-1}_{R}(x)=e^{i H_{\vec{m}} \omega(x,y)}  F(y)e^{-i H_{\vec{m}} \omega(x,y) }+8 \pi H_{\vec{m}} \delta^{2}(x-y)\;. 
\end{equation}
The extra $ \delta $-term indicates $M_{R}(x)$ is a gauge transformation singular at $ x $. When $ G=U(N) $, for $ Q $ given by (\ref{1q}), from (\ref{Q}),
\begin{equation}
[Q,M_{R}(x)]=-2 tr H_{\vec{m}} M_{R}(x)\;.
\end{equation}
$ M_{R}(x) $ carries the vortex charge $ -2 tr H_{\vec{m}} $.

Finally, for a 3d Chern-Simons-matter theory with the gauge group $ U(N)_{k}\times U(N)_{-k} $ like the ABJM, suppose $ \hat{A}_{i} $ and $ \hat{\rho} $ are the gauge field and charge density for group $ U(N)_{-k}  $, the monopole operator $ M_{[R_{1},\bar{R}_{2}]} (x) $ can be written as 
\begin{eqnarray}\label{UU}
\nonumber && M_{[R_{1},\bar{R}_{2}]}(x) \\ \nonumber &=&\exp \{\frac{ik}{2\pi} \int d^{2}y \; tr (H_{\vec{m}_{1}}[ \epsilon^{ij}A_{j}(y)a_{i}(x,y)+\frac{i}{2}\epsilon^{ij}[A_{i}(y),A_{j}(y)]\omega(x,y) +\frac{2\pi}{k}\rho(y) \omega(x,y) ])\}\\ \nonumber &&\exp \{-\frac{ik}{2\pi} \int d^{2}y \; tr (H_{\vec{m}_{2}}[ \epsilon^{ij}\hat{A}_{j}(y)a_{i}(x,y)+\frac{i}{2}\epsilon^{ij}[\hat{A}_{i}(y),\hat{A}_{j}(y)]\omega(x,y) +\frac{2\pi}{k}\hat{\rho}(y) \omega(x,y) ])\}
\;,\\
\end{eqnarray}
where $ R_{1} $ and $ R_{2} $ are two irreducible representations labeled by $ H_{\vec{m}_{1}} $ and $ H_{\vec{m}_{2}} $, respectively. The Gauss constraint in ABJM implies $ trF_{ij}=tr \hat{F}_{ij} $ \cite{tw10}. For the topological charge 
\begin{equation}\label{12qa11}
Q^{+}=\frac{1}{4\pi} \int  d^{2}x \; \epsilon^{ij}tr( F_{ij}+\hat{F}_{ij}) =\frac{1}{2\pi} \int  d^{2}x \; \epsilon^{ij}tr F_{ij} =\frac{1}{2\pi} \int  d^{2}x \; \epsilon^{ij}tr \hat{F}_{ij} \;,
\end{equation}
we have
\begin{equation}\label{12qa}
[Q^{+},M_{[R_{1},\bar{R}_{2}]}(x)]=-2tr (  H_{\vec{m}_{1}} + H_{\vec{m}_{2}})M_{[R_{1},\bar{R}_{2}]}(x)=-4tr H_{\vec{m}_{1}} M_{[R_{1},\bar{R}_{2}]}(x)=-4  tr H_{\vec{m}_{2}} M_{[R_{1},\bar{R}_{2}]}(x)
\end{equation}
with $tr H_{\vec{m}_{1}}  =tr H_{\vec{m}_{2}}   $ assumed.

\section{Monopole operators in representation $ R $}\label{sec}

The monopole operator $M_{R}(x)  $ discussed in section \ref{mr} and \ref{mr1}, although labeled by $ R $, does not transform in representation $ R $ under the gauge transformation. In the following, we will compute the gauge transformation of $  M_{R}(x)$ explicitly.

First, in 3d electromagnetic theory coupling with the matter, 
\begin{equation}
M(x)=\exp \{\frac{i}{2} \int d^{2}y \;[\Pi^{i}(y)a_{i}(x,y)  +\rho(y)\omega(x,y)   ]\}\;,
\end{equation}
which is invariant under the $ U(1) $ gauge transformation.

In 3d $ U(1) $ Chern-Simons theory coupling with the matter, 
\begin{equation}
M(x)=\exp \{\frac{ik}{4\pi} \int d^{2}y \; [ \epsilon^{ij}A_{j}(y)a_{i}(x,y) +\frac{2\pi}{k}\rho(y) \omega(x,y) ]\}\;.
\end{equation}
Under the $ U(1) $ local gauge transformation, $ A_{i} \rightarrow A_{i} + \partial_{i}\alpha $, $  \rho \rightarrow \rho$,
\begin{equation}
M(x)=e^{ik \alpha(x) }  M(x)\;. 
\end{equation}
$ M(x) $ transforms as an operator at $ x $ carrying the $ U(1) $ charge $ k $.

In 3d non-Abelian gauge theory with the group $ G $, suppose $ \mathcal{G} $ is the group composed by the local gauge transformation operator $ U $, $ \forall \; U \in   \mathcal{G} $, 
\begin{equation}
UA_{i}U^{-1}= u^{+}A_{i}u+iu^{+}\partial_{i}u\;,\;\;\;\;\;\;\;\;U\Pi_{i}U^{-1}= u^{+}\Pi_{i}u\;,\;\;\;\;\;\;\;\; U\rho U^{-1}= u^{+}\rho u   \;,
\end{equation}
where $ u \in G$ is the transformation matrix for $ U $. Concretely, for $ U(\alpha) $ given by (\ref{22w}), the related $u(y)= e^{-i\alpha(y)} $. Actions of $ U $ on $M_{R}(x)  $ in (\ref{1m}) and (\ref{2m}) are given by 
\begin{equation}\label{1t}
UM_{R}(x)U^{-1}=\exp \{i \int d^{2}y \; tr H_{\vec{m}}[u^{+}(\Pi^{i}a_{i}+i [A_{i},\Pi^{i}]\omega  + \rho \omega)u
 -u^{+} [\partial_{i}uu^{+},\Pi^{i}]u\omega    ]\}
\end{equation}
and
\begin{eqnarray}\label{2t}
\nonumber UM_{R}(x) U^{-1}&=&\exp \{\frac{ik}{2\pi} \int d^{2}y \; tr (uH_{\vec{m}}u^{+})[( \epsilon^{ij}A_{j}a_{i}+\frac{i}{2}\epsilon^{ij}[A_{i},A_{j}]\omega+\frac{2\pi}{k} \rho  \omega)\\ &+&
 i\epsilon^{ij}  [A_{i},i\partial_{j}u u^{+} ]\omega+\frac{i}{2}\epsilon^{ij} [i\partial_{i}u u^{+} ,i\partial_{j}u u^{+} ]\omega+i \epsilon^{ij}a_{i}  \partial_{j}u u^{+} ]\}\;,
\end{eqnarray}
respectively. Obviously, $ M_{R}(x)  $ does not transform in representation $ R $.

It is desirable to construct the monopole operator that would transform as a local operator in representation $ R $ under the action of $ U $. Suppose $\{ \vert \alpha \rangle \;|\alpha =1,2,\cdots,\dim R\}$ are bases for the representation $ R $, among which $ \vert 1 \rangle $ is the highest weight state. The group element $ u $ in representation $ R $ is given by $ D^{\alpha}_{\beta \; R} (u)= \langle \alpha | u | \beta \rangle$. $D^{\alpha}_{\gamma\; R} (uv)=D^{\alpha}_{\beta\; R} (u)D^{\beta}_{\gamma \;R} (v)  $. Consider 
\begin{equation}\label{1r}
\mathcal{M}^{\alpha}_{\beta\;(R,R') }(x) =\int DU \; D^{\alpha }_{\beta \;R }[u(x)]  UM_{R'}(x)U^{-1}
\end{equation}
and 
\begin{equation}\label{2r}
\mathcal{M}^{+\alpha}_{\beta\;(R,R') }(x) =\int DU \; D^{\alpha}_{\beta \;R }[u^{+}(x)]  UM_{R'}(x)U^{-1}\;,
\end{equation}
where $ R' $ is an arbitrary irreducible representation. $ \forall \; V \in   \mathcal{G} $, 
\begin{equation}\label{410q}
V\mathcal{M}^{\alpha}_{\beta\;(R,R')}(x) V^{-1}=\int DU \; D^{\alpha}_{\beta \;R }[u(x)] V UM_{R'}(x)U^{-1}V^{-1}=D^{ \alpha}_{\gamma \;R }[v^{-1}(x)] \mathcal{M}^{\gamma}_{\beta\; (R,R')}(x)
\end{equation}
\begin{equation}\label{411q}
V\mathcal{M}^{+\alpha }_{\beta\;(R,R')}(x) V^{-1}=\int DU \; D^{\alpha }_{ \beta \;R}[u^{+}(x)]  VUM_{R'}(x)U^{-1}V^{-1}
=\mathcal{M}^{+\alpha}_{\gamma \;(R,R')  }(x) D^{\gamma}_{ \beta \;R}[v(x)  ] \;.
\end{equation}
It seems that $ \mathcal{M}^{\alpha}_{\beta\;(R,R')}(x) $ with the fixed $ \beta $ transforms as a local operator in representation $ R $, while $\mathcal{M}^{+\alpha}_{\beta\;(R,R')}(x)   $ with the fixed $ \alpha$ transforms as a local operator in representation $ \bar{R} $.

However, in most cases, $  \mathcal{M}^{\alpha}_{\beta\;(R,R')}(x)$ and $\mathcal{M}^{+\alpha}_{\beta\;(R,R')}(x)    $ constructed in (\ref{410q}) and (\ref{411q}) are actually $ 0 $. Since
\begin{eqnarray}
\nonumber \mathcal{M}^{\alpha}_{\beta\;(R,R')}(x)&=&\int DU \; D^{\alpha}_{ \beta\;R}[u(x)v(x)]  UVM_{R'}(x)V^{-1}U^{-1}\\ &=&D^{ \gamma}_{ \beta\;R}[v(x)] 
\int DU \; D^{\alpha}_{ \gamma\;R}[u(x)]  UVM_{R'}(x)V^{-1}U^{-1} \;,
\end{eqnarray}
for $ V $ satisfying 
\begin{equation}\label{ivl}
VM_{R'}(x)V^{-1} = M_{R'}(x)\;,
\end{equation}
there will be 
\begin{equation}\label{iv}
\mathcal{M}^{\alpha}_{\beta\;(R,R')} (x)=\mathcal{M}^{\alpha}_{\gamma\;(R,R')} (x)D^{ \gamma}_{ \beta\;R}[v(x)]   \;,
\end{equation}
which requires 
\begin{equation}\label{ivr}
D^{ \gamma}_{ \beta\;R}[v] =\delta^{ \gamma}_{ \beta}\;,
\end{equation}
since otherwise, $ \mathcal{M}^{\alpha}_{\beta\;(R,R')} =0 $.

Let us consider the invariant group of $ M_{R}(x) $. Suppose $ G_{H} = \{e^{i H_{A}\alpha^{A}} | \;\forall \;\alpha^{A} \} $ is the Carton subgroup of $ G $. The corresponding gauge transformation operators compose $ \mathcal{G}_{H}  $ which is a subgroup of $  \mathcal{G} $. $ \forall \; V \in  \mathcal{G}_{H}$ with the related $ v=e^{i H_{A}\alpha^{A}}  $, for $ M_{R} $ in Yang-Mills-matter theory, from the gauge transformation rule (\ref{1t}), 
\begin{equation}\label{1k}
VM_{R}(x)V^{-1}=M_{R}(x) \;.
\end{equation}
On the other hand, for $ M_{R} $ in Chern-Simons-matter theory, from the gauge transformation rule (\ref{2t}), 
\begin{equation}\label{2k}
VM_{R}(x)V^{-1}=\exp \{-2i k  tr (H_{\vec{m}}  H_{A}) \alpha^{A}(x) \}M_{R}(x) \;.
\end{equation}

In Yang-Mills-matter theory, (\ref{1k}) requires $  D^{ \gamma}_{ \beta\;R}[v] =\delta^{ \gamma}_{ \beta}  $, $ \forall \; v \in  G_{H}$, which is possible only when $ R $ is the identity representation $ I $, in which case, 
\begin{equation}
\mathcal{M}_{(I,R)}(x) =\int DU \; UM_{R}(x)U^{-1}\;.
\end{equation}
So the only non-zero $ \mathcal{M} $ that can be constructed is $\mathcal{M}_{(I,R)}(x)  $ which is gauge invariant.

In Chern-Simons-matter theory, $ M_{R} $ transforms as (\ref{2k}) under the action of $ \mathcal{G}_{H}  $. When $ G=U(N) $, for $ R' $ labeled by 
\begin{equation}
H_{\vec{m}'}    =diag(\frac{m'_{1}}{2},\cdots,\frac{m'_{N}}{2}) \;,
\end{equation}
consider $ N $ $ U(1) $ transformation groups 
 \begin{equation}
v_{i}=diag(\underbrace{1,\cdots,1}_{i-1},e^{i\theta_{i}},\underbrace{1,\cdots,1}_{N-i}) \in G_{H}\;,\;\;\;\;\;\;\;\;i=1,\cdots,N\;,
\end{equation}
from (\ref{2k}), the corresponding $V_{i}  $ will make
\begin{equation}
V_{i}M_{R'}(x)V_{i}^{-1}=\exp \{-i k  m'_{i} \theta_{i}(x) \}M_{R'}(x) \;.
\end{equation}
Accordingly, for $ \mathcal{M}^{\alpha}_{\beta\;(R,R')} (x)\neq 0 $, $ \mathcal{M}^{\alpha}_{\beta\;(R,R')} (x)  $ should satisfy
\begin{equation}
\mathcal{M}^{\alpha}_{\gamma\;(R,R')} (x)D^{ \gamma}_{ \beta\;R}[v_{i}(x)]=\exp \{i k m'_{i} \theta_{i}(x)  \}\mathcal{M}^{\alpha}_{\beta\;(R,R')} (x)  \;.
\end{equation}
So, in addition to the condition (\ref{ivr}) for $ V $ satisfying (\ref{ivl}), $ M_{R'} $ and $  \mathcal{M}^{\alpha}_{\beta\;(R,R')} $ should also have the opposite $ U(1) $ charges, which is possible only when $ R$ is the representation labeled by $  kH_{\vec{m}'} $ and $\vert \beta \rangle$ is the highest weight state in representation $ R $.

For example, when
 \begin{equation}
H_{\vec{m}'}=diag(\frac{m'_{1}}{2},\underbrace{0,\cdots,0}_{N-1})\;,
\end{equation}
the invariant group of $  M_{R'} $ and $ \mathcal{M}^{\alpha}_{\beta\;(R,R')}   $ is $ U(N-1) $, and $ \mathcal{M}^{\alpha}_{\beta\;(R,R')}   $ should satisfy
\begin{equation}
\mathcal{M}^{\alpha}_{\gamma\;(R,R')} (x)D^{ \gamma}_{ \beta\;R}[v_{1}(x)]=\exp \{i k m'_{1} \theta_{1}(x)  \}\mathcal{M}^{\alpha}_{\beta\;(R,R')} (x)  \;.
\end{equation}
$ R $ must be the $ km'_{1} $-symmetric representation with $ \vert \beta \rangle $ the highest weight state 
\begin{equation}
\bigotimes_{km'_{1}} ( 1,\underbrace{0,\cdots,0}_{N-1})\;.
\end{equation}
When 
 \begin{equation}
H_{\vec{m}'}=diag(\frac{1}{2},\frac{1}{2},\underbrace{0,\cdots,0}_{N-2})\;,
\end{equation}
the invariant group of $  M_{R'} $ and $ \mathcal{M}^{\alpha}_{\beta\;(R,R')}   $ is $ SU(2)\times U(N-2) $, and $ \mathcal{M}^{\alpha}_{\beta\;(R,R')}   $ should satisfy
\begin{equation}
\mathcal{M}^{\alpha}_{\gamma\;(R,R')} (x)D^{ \gamma}_{ \beta\;R}[v_{i}(x)]=\exp \{i k  \theta_{i}(x)  \} \mathcal{M}^{\alpha}_{\beta\;(R,R')} (x)  \;,\;\;\;\;\;\;\;\;i=1,2\;.
\end{equation}
$ R $ must be an irreducible representation with the Young tableau containing $ k $ boxes in the first two rows. $ \vert \beta \rangle $ is the highest weight state
\begin{equation}
\bigotimes_{k} \left[ ( 1,\underbrace{0,\cdots,0}_{N-1}) \bigotimes ( 0,1,\underbrace{0,\cdots,0}_{N-2})  - ( 0,1,\underbrace{0,\cdots,0}_{N-2}) \bigotimes ( 1,\underbrace{0,\cdots,0}_{N-1})  \right]  \;.
\end{equation}

To conclude, in Chern-Simons-matter theory with the level $ k $, the monopole operator in representation $ R $ labeled by $ H_{\vec{m}} $ can be constructed as 
\begin{equation}\label{328h}
\mathcal{M}^{\alpha}_{R }(x) =\frac{1}{\sqrt{\mathcal{N}_{R}}}\int_{U \in \mathcal{G}} DU \; D^{\alpha }_{1 \;R }[u(x)]  UM_{R/k}(x)U^{-1}\;,
\end{equation}
where $ \vert 1 \rangle $ is the highest weight state in presentation $ R $ and $ M_{R/k}(x) $ is the monopole operator labeled by $ H_{\vec{m}/k}   $. Since $ m_{i} /k$ must be integers, for $ k>1 $, not all of $ R $ can be realized. For example, if $ R $ is the fundamental representation, $ k $ must be $ 1 $; if $ R $ is the 2-symmetric representation, $ k $ can only be $ 1 $ or $ 2 $. The integration measure $ DU $ is normalized with 
\begin{equation}\label{328hb}
\int_{U \in \mathcal{G}} DU=1 \;.
\end{equation}

We also add a divided factor $\sqrt{\mathcal{N}_{R}}  $. Suppose $\mathcal{G}_{S}  \subset  \mathcal{G}$ is the stationary group of the integrand, $ \mathcal{N}_{R} $ is given by 
\begin{equation}
\mathcal{N}_{R} =\int_{U \in \mathcal{G}_{S}} DU   \;.
\end{equation}
By stationary group, we mean $ \forall \; V \in   \mathcal{G}_{S}$, 
\begin{equation}
D^{\alpha }_{1 \;R }[u(x)v(x)]  UVM_{R/k}(x)V^{-1}U^{-1}=D^{\alpha }_{1 \;R }[u(x)] UM_{R/k}(x)U^{-1}\;,\;\;\;\;\;\;\;\; \forall \; U \in   \mathcal{G} \;,
\end{equation}
which is equivalent to the condition 
\begin{equation}\label{sta}
D^{\beta}_{1\;R}[v(x)] VM_{R/k}(x)V^{-1}=\delta^{ \beta}_{ 1}M_{R/k}(x)\;.
\end{equation}
The associated $ v $ could also compose a stationary group $ G_{s} $, under which $ \vert 1 \rangle $ is invariant up to a phase. $\forall \; v  \in G_{s} $, $v \vert 1\rangle = e^{i\theta}  \vert 1\rangle$, 
\begin{equation}
D^{\beta}_{1\;R}[v(x)] = \langle \beta \vert v(x) \vert 1\rangle =e^{i\theta(x)} \delta^{\beta}_{1}\;,\;\;\;\;\;\;\;\;\;
VM_{R/k}(x)V^{-1}=e^{-i\theta(x)}M_{R/k}\;.
\end{equation}

Similarly, the monopole operator in representation $ \bar{R} $ is
\begin{equation}\label{434g}
\mathcal{M}_{\alpha R }(x) =\frac{1}{\sqrt{\mathcal{N}_{R}}}\int_{U \in \mathcal{G}}  DU \; D^{1}_{\alpha \;R }[u^{+}(x)]  UM^{-1}_{R/k}(x)U^{-1}\;,
\end{equation}
$ D^{\alpha}_{\beta\;R}(u^{+}) =D^{\beta*}_{\alpha\;R}(u) $. $ \mathcal{M}^{+}_{\alpha R }(x) = \mathcal{M}^{\alpha}_{R }(x)$. Under the local gauge transformation $ V $, 
\begin{equation}
V\mathcal{M}^{\alpha}_{R}(x) V^{-1}=D^{\alpha}_{\beta\;R}[v^{-1}(x)] \mathcal{M}^{\beta}_{R }(x)\;,\;\;\;\;\;\;\;\;\;
V\mathcal{M}_{\alpha R }(x) V^{-1}
=\mathcal{M}_{\beta R }(x) D^{\beta}_{ \alpha\;R}[v(x)  ] \;.
\end{equation}
If $ O^{\alpha}_{R}(x) $ and $ O_{\alpha R }(x) $ are normal local operators in representation $ R $ and $ \bar{R} $ with 
\begin{equation}\label{328ha}
VO^{\alpha}_{R}(x) V^{-1}=D^{\alpha}_{\beta\;R}[v^{-1}(x)] O^{\beta}_{R }(x)\;,\;\;\;\;\;\;\;\;\;
VO_{\alpha R }(x) V^{-1}
=O_{\beta R }(x) D^{\beta}_{ \alpha\;R}[v(x)  ] \;,
\end{equation}
$ \mathcal{M}^{\alpha}_{R}(x) O_{\alpha R }(x)$ and $ \mathcal{M}_{\alpha R }(x)    O^{\alpha}_{R}(x) $ will be gauge invariant. In this way, the vortex-charged gauge invariant operators can be constructed.

Recall that in (\ref{24f}), $ a_{i} $ can differ by a local gauge transformation. With $ a_{i} \rightarrow  a_{i}+\partial_{i} \sigma $, $ M_{R}(x)  \rightarrow U(H_{\vec{m}}\sigma) M_{R}(x)   $  up to a phase. The different $ a_{i}  $ will also give rise to the different $\mathcal{M}_{R}  $. Such ambiguity can be eliminated by Gauss constraint, which is imposed via a projection to the physical Hilbert space $  \mathcal{H}_{phy}  $. $ \forall \; \vert \psi \rangle  \in \mathcal{H}_{phy} $, $\forall \; U \in \mathcal{G}  $, $ U \vert \psi\rangle = \vert \psi \rangle$. From (\ref{328h}), (\ref{328ha}) and (\ref{328hb}), the action of the physical operator $ \mathcal{M}^{\alpha}_{R }(x)  O_{\alpha R }(x) $ on physical state $ \vert  \psi_{1}\rangle  $ is
\begin{eqnarray}
\nonumber \mathcal{M}^{\alpha}_{R }(x)  O_{\alpha R }(x) \vert  \psi_{1}\rangle  &=&\frac{1}{\sqrt{\mathcal{N}_{R}}}\int_{U \in \mathcal{G}} DU \;  UM_{R/k}(x)O_{1 R }(x) \vert  \psi_{1}\rangle\\ &=&\frac{1}{\sqrt{\mathcal{N}_{R}}}\int_{U \in \mathcal{G}} DU \;  UU(H_{\vec{m}/k}\sigma)  M_{R/k}(x)O_{1 R }(x) \vert  \psi_{1}\rangle  \in  \mathcal{H}_{phy}  \;,
\end{eqnarray}
and the projection of $  \mathcal{M}^{\alpha}_{R }(x)  O_{\alpha R }(x)  $ in $  \mathcal{H}_{phy}   $ is
\begin{eqnarray}
 \label{438a}\langle  \psi_{1} \vert \mathcal{M}^{\alpha}_{R }(x) O_{\alpha R }(x) \vert  \psi_{2}\rangle  &=&\frac{1}{\sqrt{\mathcal{N}_{R}}}  \langle  \psi_{1} \vert  M_{R/k}(x)O_{1 R }(x) \vert  \psi_{2}\rangle\\ &=&\frac{1}{\sqrt{\mathcal{N}_{R}}}  \langle  \psi_{1} \vert U(H_{\vec{m}/k}\sigma)  M_{R/k}(x)O_{1 R }(x) \vert  \psi_{2}\rangle\;. 
\end{eqnarray}
So with $ a_{i} $ replaced by $  a_{i}+\partial_{i} \sigma $, the physical state produced by $ \mathcal{M}^{\alpha}_{R }(x)  O_{\alpha R }(x) $ and the matrix elements of $ \mathcal{M}^{\alpha}_{R }(x)  O_{\alpha R }(x) $ in $ \mathcal{H}_{phy} $ remain the same. We may select an arbitrary $  a_{i} $ to construct $M_{R}  $, and the obtained physical operators in $  \mathcal{H}_{phy}$ are identical. In fact, when computing the matrix element in $ \mathcal{H}_{phy} $, (\ref{438a}) can be equivalently interpreted in path integral language. From (\ref{a77}), the action of $  M_{R/k}(x) $ produces a vortex-charged configuration, which is then combined with $ O_{1 R }(x) $ to compensate the $ U(1) $ charges. The integration over gauge equivalent configurations gives the matrix element in physical Hilbert space.

In Chern-Simons-matter theory with the gauge group $ U(N)_{k}\times U(N)_{-k} $, $ M_{[R_{1},\bar{R}_{2}]}(x) $ is given by (\ref{UU}), and the corresponding monopole operator in representation $ (R_{1},\bar{R}_{2} ) $ can be constructed as 
\begin{eqnarray}\label{435}
\nonumber \mathcal{M}_{\hat{\alpha}(R_{1},\bar{R}_{2} )  }^{\alpha}(x)  &=&\frac{1}{\sqrt{\mathcal{N}_{R_{1}}\mathcal{N}_{R_{2}}}}  \int_{U \in \mathcal{G}}  DU\int_{\hat{U} \in \hat{\mathcal{G}}} D\hat{U} \; D^{\alpha }_{1 \;R_{1} }[u(x)] D^{1}_{\hat{\alpha} \;R_{2}  }[\hat{u}^{+}(x)]     U\hat{U}M_{[R_{1},\bar{R}_{2}]/k}(x)U^{-1}\hat{U}^{-1}\\ &=& \mathcal{M}^{\alpha}_{R_{1} }(x)\mathcal{M}_{\hat{\alpha} R_{2} }(x)\;,
\end{eqnarray}
where $ \mathcal{M}^{\alpha}_{R_{1} }(x)$ and $\mathcal{M}_{\hat{\alpha} R_{2} }(x) $ are monopole operators for the first and second $ U(N) $ groups. Under the local gauge transformation $  V\hat{V}$, 
\begin{equation}
V\hat{V}\mathcal{M}_{\hat{\alpha}(R_{1},\bar{R}_{2} )  }^{\alpha}(x)  V^{-1}\hat{V}^{-1}=D^{\alpha}_{\beta\;R_{1}}[v^{-1}(x)]  \mathcal{M}_{\hat{\beta}(R_{1},\bar{R}_{2} )  }^{\beta}(x) D^{\hat{\beta}}_{ \hat{\alpha}\;R_{2}}[\hat{v}(x)  ]\;.
\end{equation}
The topological charge $ Q^{+} $ is gauge invariant, so from (\ref{12qa}) and (\ref{435}), 
\begin{equation}
[Q^{+},\mathcal{M}_{\hat{\alpha}(R_{1},\bar{R}_{2} )   }^{\alpha}(x)  ]=-4 tr H_{\vec{m}_{1}/k} \mathcal{M}_{\hat{\alpha}(R_{1},\bar{R}_{2} )   }^{\alpha}(x) =-4 tr H_{\vec{m}_{2}/k} \mathcal{M}_{\hat{\alpha}(R_{1},\bar{R}_{2} )   }^{\alpha}(x)  \;.
\end{equation}
When $ R_{1} =R_{2}=R$, $ \mathcal{M}_{\hat{\alpha}(R,\bar{R} )  }^{\alpha}(x)  $ can be denoted as $ \mathcal{M}_{\hat{\alpha}R  }^{\alpha}(x)  $.
\begin{equation}
\mathcal{M}_{\hat{\alpha}R  }^{\alpha}(x)  =\mathcal{M}^{\alpha}_{R }(x)\mathcal{M}_{\hat{\alpha} R }(x)
\end{equation}
is in the $ (R,\bar{R}) $ representation of $ U(N)\times U(N) $. $\mathcal{M}_{\hat{\alpha}R  }^{\alpha +}(x) =\mathcal{M}_{\alpha R  }^{\hat{\alpha}}(x)   $.

In ABJM theory, monopole operators with $  R_{1} =R_{2}$ are used to construct gauge invariant chiral operators. The generic monopole operators with $  R_{1} \neq R_{2} $ should also exist and are required to make the large $ N $ spectrum of the protected operators agree with that of the gravity states \cite{tw10a}.

\section{The contraction relations for monopole operators}\label{cont}

In this section, we will prove some contraction relations for monopole operators $ \mathcal{M}^{\alpha}_{R }(x)$ and $\mathcal{M}_{\alpha R }(x) $ that will be used in section \ref{SO8}.

Consider the Chern-Simons-matter theory with the gauge group $ U(N) $ and the level $ k=1 $. If $ R $ is the fundamental representation $\mathbf{N}$, the related $ H_{\vec{m}} $ is
\begin{equation}\label{h1}
H_{[1]}=diag(\frac{1}{2},\underbrace{0,\cdots,0}_{N-1})\;.
\end{equation}
The corresponding monopole operator is 
\begin{equation}
\mathcal{M}^{a }(x) =\frac{1}{\sqrt{\mathcal{N}_{[1]}}}\int_{U \in \mathcal{G}}  DU \; D^{a}_{ 1}[u(x)]  UM_{[1]} (x)U^{-1}\;,
\end{equation}
where $ a=1,\cdots,N $, $ D^{a}_{ b}[u] $ is the fundamental representation of $ u \in U(N) $, and $ M_{[1]} (x) $ is $ M_{R} (x)  $ with $ H_{\vec{m}}   $ given by (\ref{h1}). 
\begin{equation}
\mathcal{M}_{a}(x) =\frac{1}{\sqrt{\mathcal{N}_{[1]}}}\int_{U \in \mathcal{G}}  DU \; D_{a}^{1}[u^{+}(x)]   UM_{[1]}^{-1} (x)U^{-1}
\end{equation}
is the monopole operator in antifundamental representation $\bar{\mathbf{N}}  $. $\mathcal{M}^{+}_{a}(x) = \mathcal{M}^{a }(x)  $. Under the local gauge transformation, 
\begin{equation}
V\mathcal{M}^{a}(x) V^{-1}=D^{a}_{b}[v^{-1}(x)] \mathcal{M}^{b}(x)\;,\;\;\;\;\;\;\;\;\;
V\mathcal{M}_{a }(x) V^{-1}
=\mathcal{M}_{b}(x) D^{b}_{ a}[v(x)  ] \;.
\end{equation}

The contraction of $\mathcal{M}_{a}(x)$ and $\mathcal{M}^{a }(x)  $ is
\begin{eqnarray}
\nonumber \mathcal{M}_{a}(x)\mathcal{M}^{a }(x) &=&\frac{1}{\mathcal{N}_{[1]}}\int_{U,V \in \mathcal{G}} DV DU  \;  D_{a}^{1}[v^{+}(x)] D^{a}_{ 1}[u(x)]  VM_{[1]}^{-1} (x)V^{-1}   UM_{[1]} (x)U^{-1}\\  &=&\frac{1}{\mathcal{N}_{[1]}}\int_{U \in \mathcal{G}}   DU \;  D_{1}^{1}[ u(x)] \int_{V \in \mathcal{G}}  DV\; V[M_{[1]}^{-1} (x)   UM_{[1]} (x)U^{-1}]V^{-1}\;. 
\end{eqnarray}
$ \forall \; W \in \mathcal{G} $,
\begin{equation}\label{65}
\int_{V \in \mathcal{G}} DV\; V[M_{[1]}^{-1} (x)   UM_{[1]} (x)U^{-1}]V^{-1}=\int_{V \in \mathcal{G}} DV\; VW[M_{[1]}^{-1} (x)   UM_{[1]} (x)U^{-1}]W^{-1}V^{-1}\;.
\end{equation}
For the given $ U $, one can always find some $ W $ with
\begin{equation}
W[M_{[1]}^{-1} (x)   UM_{[1]} (x)U^{-1}]W^{-1} =e^{i\theta} [M_{[1]}^{-1} (x)   UM_{[1]} (x)U^{-1}]\;.
\end{equation}
As a result, (\ref{65}) $ \neq 0 $ only when $ U \in \mathcal{G}_{S} $, $ u \in G_{s}=U(1)\times U(N-1) $, and then 
\begin{equation}
 D_{1}^{1}[ u(x)] UM_{[1]} (x)U^{-1}=M_{[1]} (x) \;,
\end{equation}
\begin{equation}
\mathcal{M}_{a}(x)\mathcal{M}^{a }(x)=\frac{1}{\mathcal{N}_{[1]}}\int_{U \in \mathcal{G}_{S}} DU\int_{V \in \mathcal{G} } DV =1\;.
\end{equation}
$\mathcal{M}^{a}  $ and $\mathcal{M}_{a}  $ in representations $\mathbf{N}  $ and $ \bar{\mathbf{N}}  $ do not exist when $ k >1 $.

When $ k=1 $, the 2-symmetric representation $\mathbf{N}^{2}_{sym}  $ is labeled by 
 \begin{equation}\label{h2}
H_{[2]}=diag(1,\underbrace{0,\cdots,0}_{N-1})\;.
\end{equation}
The corresponding monopole operator is 
\begin{equation}
\mathcal{M}^{ab }(x) =\frac{1}{\sqrt{\mathcal{N}_{[2]}}}\int_{U \in \mathcal{G}} DU \; D^{a}_{ 1}[u(x)]D^{b}_{ 1}[u(x)]   UM_{[2]} (x)U^{-1}\;,
\end{equation}
where $  M_{[2]} (x) $ is $ M_{R} (x)  $ with $ R $ labeled by (\ref{h2}). The monopole operator in $\bar{\mathbf{N}}^{2}_{sym}  $ representation is
\begin{equation}
\mathcal{M}_{ab}(x) =\frac{1}{\sqrt{\mathcal{N}_{[2]}}}\int_{U \in \mathcal{G}} DU \; D_{a}^{1}[u^{+}(x)]D_{b}^{1 }[u^{+}(x)]   UM_{[2]}^{-1} (x)U^{-1}\;.
\end{equation}
$\mathcal{M}^{+}_{ab}(x) =  \mathcal{M}^{ab }(x)$. Under the local gauge transformation, 
\begin{equation}
V\mathcal{M}^{ab}(x) V^{-1}=D^{a}_{c}[v^{-1}(x)] D^{b}_{d}[v^{-1}(x)] \mathcal{M}^{cd}(x)\;,\;\;\;\;\;\;\;
V\mathcal{M}_{ab }(x) V^{-1}
=\mathcal{M}_{cd}(x) D^{c}_{ a}[v(x)  ]  D^{d}_{ b}[v(x)  ]  \;.
\end{equation}

The contraction of $\mathcal{M}_{ab}(x)$ and $ \mathcal{M}^{bc }(x)  $ is
\begin{eqnarray}
\nonumber && \mathcal{M}_{ab}(x) \mathcal{M}^{bc }(x)\\\nonumber &=&\frac{1}{\mathcal{N}_{[2]}}\int_{V,U \in \mathcal{G}} DV DU \; D_{a}^{1}[v^{+}(x)]D_{b}^{1 }[v^{+}(x)]    D^{b}_{ 1}[u(x)]D^{c}_{ 1}[u(x)] VM_{[2]}^{-1} (x)V^{-1}    UM_{[2]} (x)U^{-1}\\ \nonumber &=&\frac{1}{\mathcal{N}_{[2]}}\int_{U \in \mathcal{G}} DU\;  D^{1}_{ 1}[u(x)]D^{b}_{ 1}[u(x)] \int_{V \in \mathcal{G}} DV  \; D_{a}^{1}[v^{+}(x)] D^{c}_{ b}[v(x)] V[M_{[2]}^{-1} (x)  UM_{[2]} (x)U^{-1}]V^{-1}\;. \\
\end{eqnarray}
$ \forall \; W \in  \mathcal{G}$, 
\begin{eqnarray}\label{611}
\nonumber && \int_{V \in \mathcal{G}} DV  \; D_{a}^{1}[v^{+}(x)] D^{c}_{ b}[v(x)] V[M_{[2]}^{-1} (x)  UM_{[2]} (x)U^{-1}]V^{-1} \\&=&\int_{V \in \mathcal{G}} DV  \; D_{a}^{1}[w^{+}(x)v^{+}(x)] D^{c}_{ b}[v(x)w(x)] V W[M_{[2]}^{-1} (x)  UM_{[2]} (x)U^{-1}]W^{-1}V^{-1}\;.
\end{eqnarray}
For the given $ U $, one can always find some $ W $ with 
\begin{equation}
W[M_{[2]}^{-1}  UM_{[2]} U^{-1}]W^{-1}=e^{i\theta} [M_{[2]}^{-1}   UM_{[2]} U^{-1}]\;.
\end{equation}
(\ref{611}) $ \neq 0 $ only when 
\begin{equation}
 D_{a}^{1}[w^{+}v^{+}] D^{c}_{ b}[vw]=e^{-i\theta}  D_{a}^{1}[v^{+}] D^{c}_{ b}[v]\;,\;\;\;\;\;\;\;\forall \; v \in G\;.
\end{equation}
We should at least have 
 \begin{equation}
u H_{[2]}u^{+}=diag(\underbrace{0,\cdots,0}_{m},1, \underbrace{0,\cdots,0}_{N-m-1})\;.
\end{equation}
Moreover, $ m $ must be $ 0 $, since otherwise, $D^{1}_{ 1}[u]  =0$. So we have $ U \in \mathcal{G}_{s} $, $ u H_{[2]}u^{+}= H_{[2]}$, in which case, $ D^{b}_{ 1}[u]\neq 0$ only when $b= 1  $, $ (D^{1}_{ 1}[u(x)])^{2} UM_{[2]} (x)U^{-1} = M_{[2]}$. 
\begin{eqnarray}\label{520}
\nonumber  \mathcal{M}_{ab}(x) \mathcal{M}^{bc }(x)&=&\frac{1}{\mathcal{N}_{[2]}}\int_{U \in \mathcal{G}_{S}} DU\;  \int_{V \in \mathcal{G}} DV  \; D_{a}^{1}[v^{+}(x)] D^{c}_{ 1}[v(x)] =\int_{v \in G} dv  \; D_{a}^{1}[v^{+}(x)] D^{c}_{ 1}[v(x)]\\&=&\frac{\delta^{c}_{a}}{N}\;,
\end{eqnarray}
where we have used 
\begin{equation}
\int_{u \in G} du \;D^{\alpha}_{\beta \;R}[u^{+}]D^{\rho}_{\sigma \;R}[u]=\frac{1}{d_{R}}\delta^{\rho}_{\beta}\delta^{\alpha}_{\sigma}\;,
\end{equation}
$\int_{u \in G} du=1  $. Again, $\mathcal{M}_{ab}(x) \mathcal{M}^{ba }(x)= 1 $.

The same conclusion holds for $ k=2 $ with $H_{[2]}  $ in (\ref{h2}) replaced by $H_{[1]}  $ in (\ref{h1}). When $ k>2 $, $ \mathcal{M}^{ab}(x)  $ and $ \mathcal{M}_{ab}(x)  $ in representations $\mathbf{N}^{2}_{sym}  $ and $\bar{\mathbf{N}}^{2}_{sym}  $ do not exist. When $ k=1 $, with $ H_{\vec{m}} $ taken to be 
 \begin{equation}
H_{[1,1]}=diag(\frac{1}{2},\frac{1}{2},\underbrace{0,\cdots,0}_{N-2})\;,
\end{equation}
we can also get $ \mathcal{M}'^{ab}(x)  $ in 2-antisymmetric representation $\mathbf{N}^{2}_{asym}  $ and $ \mathcal{M}'_{ab}(x)  $ in conjugate representation $\bar{\mathbf{N}}^{2}_{asym}  $.

In Chern-Simons-matter theory with the gauge group $ U(N)_{1}\times U(N)_{-1} $, the monopole operator in representation $ (\mathbf{N},\bar{\mathbf{N}}) $ is 
 \begin{equation}
\mathcal{M}_{\hat{a}  }^{a}(x) =\frac{1}{N_{[1]}}\int DUD\hat{U} \; D^{a }_{1  }[u(x)] D^{1}_{\hat{a}  }[\hat{u}^{+}(x)]     U\hat{U}M_{[1,\bar{1}]}(x)U^{-1}\hat{U}^{-1}=\mathcal{M}^{a}(x)\mathcal{M}_{\hat{a} }(x)\;,
\end{equation}
where $ M_{[1,\bar{1}]}(x) $ is given by (\ref{UU}) with $ k=1 $ and $ H_{\vec{m}}=H_{[1]} $, $ a,\hat{a} =1,\cdots,N$. $ \mathcal{M}_{\hat{a}  }^{a+}(x) =\mathcal{M}_{a }^{\hat{a} }(x)  $. Under the local gauge transformation, 
\begin{equation}
 V\hat{V}\mathcal{M}_{\hat{a}  }^{a}(x) V^{-1}\hat{V}^{-1}=D^{a}_{b}[v^{-1}(x)] \mathcal{M}_{\hat{b}  }^{b}(x)  D^{\hat{b} }_{ \hat{a} }[\hat{v}(x)  ]
\;.
\end{equation}
The monopole operators in $(\mathbf{N}^{2}_{sym}, \bar{\mathbf{N}}^{2}_{sym} ) $ and $ ( \bar{\mathbf{N}}^{2}_{sym} ,\mathbf{N}^{2}_{sym})  $ representations are
 \begin{equation}
\mathcal{M}_{\hat{a} \hat{b} }^{ab}(x) =\mathcal{M}^{ab}(x)   \mathcal{M}_{\hat{a} \hat{b} }(x)     \;,\;\;\;\;\;\;\;\;\;
\mathcal{M}^{\hat{a} \hat{b} }_{ab}(x) =\mathcal{M}_{ab}(x)   \mathcal{M}^{\hat{a} \hat{b} }(x)     \;.
\end{equation}
$\mathcal{M}_{\hat{a} \hat{b} }^{ab+}(x) = \mathcal{M}^{\hat{a} \hat{b} }_{ab}(x)  $. From (\ref{520}), 
\begin{equation}\label{521}
\mathcal{M}_{\hat{a} \hat{b} }^{ab}(x)\mathcal{M}^{\hat{a} \hat{c} }_{ac}(x) =\mathcal{M}^{ab}(x)   \mathcal{M}_{\hat{a} \hat{b} }(x)    \mathcal{M}_{ac}(x)   \mathcal{M}^{\hat{a} \hat{c} }(x) =\frac{1}{N^{2}} \delta^{b}_{c}  \delta^{\hat{c} }_{\hat{b} }\;.
\end{equation}
With a rescaling 
\begin{equation}
 \mathcal{M}_{\hat{a} \hat{b} }^{ab}(x)  \rightarrow  N\mathcal{M}_{\hat{a} \hat{b} }^{ab}(x) \;,\;\;\;\;\;\;\mathcal{M}^{\hat{a} \hat{c} }_{ac}(x)  \rightarrow N\mathcal{M}^{\hat{a} \hat{c} }_{ac}(x)  \;,
\end{equation}
we get 
\begin{equation}\label{427f}
\mathcal{M}_{\hat{a} \hat{b} }^{ab}(x)\mathcal{M}^{\hat{a} \hat{c} }_{ac}(x) =\delta^{b}_{c}  \delta^{\hat{c} }_{\hat{b} }
\end{equation}
which is the contraction relation needed in section \ref{SO8}.

When $ k=1 $, except for $ \mathcal{M}_{\hat{a} \hat{b} }^{ab}(x)  $ and $ \mathcal{M}^{\hat{a} \hat{b} }_{ab}(x)  $, there are $ 6 $ additional monopole operators carrying the vortex charge $ \pm 4 $ :
 \begin{equation}\label{sasq}
\mathcal{M}'^{ab}_{\hat{a} \hat{b} }(x) =\mathcal{M}'^{ab}(x)   \mathcal{M}'_{\hat{a} \hat{b} }(x)     \;,\;\;\;\;\;\;\;\;\;
\mathcal{M}'^{\hat{a} \hat{b} }_{ab}(x) =\mathcal{M}'_{ab}(x)   \mathcal{M}'^{\hat{a} \hat{b} }(x)     
\end{equation}
in $(\mathbf{N}^{2}_{asym}, \bar{\mathbf{N}}^{2}_{asym} ) $ and $ ( \bar{\mathbf{N}}^{2}_{asym} ,\mathbf{N}^{2}_{asym})  $ representations, 
\begin{equation}\label{SAS}
 \mathcal{M}''^{ab}_{\hat{a}\hat{b}} (x) =\mathcal{M}^{ab}(x) \mathcal{M}'_{\hat{a}\hat{b}} (x) \;,\;\;\;\;\;\;\;\;\;  \mathcal{M}''^{\hat{a}\hat{b}}_{ab} (x) =\mathcal{M}_{ab}(x) \mathcal{M}'^{\hat{a}\hat{b}} (x)
\end{equation}
in $ (\mathbf{N}^{2}_{sym},\bar{\mathbf{N}}^{2}_{asym}) $ and $ (\bar{\mathbf{N}}^{2}_{sym},\mathbf{N}^{2}_{asym})   $ representations and 
\begin{equation}\label{SAS1}
 \mathcal{M}'''^{ab}_{\hat{a}\hat{b}} (x) =\mathcal{M}'^{ab} (x)\mathcal{M}_{\hat{a}\hat{b}} (x) \;,\;\;\;\;\;\;\;\;\;  \mathcal{M}'''^{\hat{a}\hat{b}}_{ab}(x)  =\mathcal{M}'_{ab} (x)\mathcal{M}^{\hat{a}\hat{b}} (x)
\end{equation}
in $ (\mathbf{N}^{2}_{asym},\bar{\mathbf{N}}^{2}_{sym})$ and $  (\bar{\mathbf{N}}^{2}_{asym},\mathbf{N}^{2}_{sym}) $ representations. As will be discussed in section \ref{SO8}, in ABJM theory, such operators can be used to construct the superconformal current multiplets.

\section{Classical conformal dimension of monopole operators}\label{cla}

In this section, we will compute the classical conformal dimension of the monopole operator from the action of the dilation $ D $. Unlike \cite{Hoo1}, the calculation is entirely carried out at the kinetical level without the knowledge of the exact Lagrangian. The classical conformal dimension may receive quantum corrections from the interactions unless there are supersymmetries to protect it.

In (d+1)-dimensional conformal field theory, for a field $ \phi $ with the conformal dimension $ \Delta $, the conjugate momentum $ \pi $ should have the conformal dimension $ \Delta'=d- \Delta $. The action of the dilatation operator $ D $ is 
\begin{equation}
-i [D, \phi]= \Delta \phi + x^{\mu}\partial_{\mu}\phi\;,\;\;\;\;\;\;\;
-i [D, \pi]=(d- \Delta )\pi  + x^{\mu}\partial_{\mu}\pi \;.
\end{equation}
$ \mu = 0,1,\cdots,d $. The charge density $ \rho $ has the conformal dimension $  d$ with
\begin{equation}
-i [D, \rho]=d \rho+ x^{\mu}\partial_{\mu}\rho \;.
\end{equation}

We will only consider the 3d Chern-Simons-matter theory which can be classically conformal invariant. The gauge field $ A $ and its conjugate momentum both have the conformal dimension $ 1 $, in consistence with the commutation relation (\ref{23c}). The charge density $ \rho $ has the conformal dimension $  2$. At $ x^{0}=0 $, 
\begin{equation}
-i [D, A_{i}]= A_{i} + x^{j}\partial_{j}A_{i}\;,\;\;\;\;\;\;\;-i [D, \rho]=2 \rho+ x^{i}\partial_{i}\rho \;,
\end{equation}
where $ i,j=1,2 $. The monopole operator $  M_{R}(x)  $ is given by (\ref{2m}). Let $ M_{R}(x)  =\exp \{i m_{R}(x)\}  $, where
\begin{equation}
m_{R}(x)=\frac{k}{2\pi} \int d^{2}y \; tr H_{\vec{m}}\{ \epsilon^{ij}A_{j}(y)a_{i}(x,y)+[\frac{i}{2}\epsilon^{ij}[A_{i}(y),A_{j}(y)] +\frac{2\pi}{k}\rho(y)  ]\omega(x,y)\}\;.
\end{equation}
$a_{i}(x,y)  $ is taken to be 
\begin{equation}\label{55f}
a_{i}(x,y)=2\epsilon_{ji}\frac{x^{j}-y^{j}}{|x-y|^{2}}\;.
\end{equation}
\begin{eqnarray}
\nonumber && -i[D,m_{R}(x)] \\ \nonumber &=&\frac{k}{2\pi} \int d^{2}y \; tr H_{\vec{m}}\{ \epsilon^{ij}[A_{j}(y) + y^{k}\partial^{y}_{k}A_{j}(y)]a_{i}(x,y)+2[\frac{i}{2}\epsilon^{ij}[A_{i}(y),A_{j}(y)] +\frac{2\pi}{k}\rho(y)  ]\omega(x,y)\\ \nonumber &+&  y^{k}\partial^{y}_{k}[\frac{i}{2}\epsilon^{ij}[A_{i}(y),A_{j}(y)] +\frac{2\pi}{k}\rho(y)  ]\omega(x,y)\}\\ &=&x^{k}\partial_{k}m_{R}(x)\;,
\end{eqnarray}
where we have used 
\begin{equation}\label{38u}
(x^{j}-y^{j})\partial^{y}_{j}a_{i}=a_{i}\;,\;\;\;(x^{i}-y^{i})a_{i}=0\;,\;\;\; \partial^{y}_{k}  a_{i}(x,y)=- \partial^{x}_{k}  a_{i}(x,y)\;,\;\;\; \partial^{y}_{k}   \omega(x,y)=-  \partial^{x}_{k} \omega(x,y)  \;.
\end{equation}
Therefore,
\begin{equation}\label{68h}
 -i[D,M_{R}(x)]=x^{i}\partial_{i}M_{R}(x)\;.
\end{equation}
The classical conformal dimension of $ M_{R}(x) $ is $ 0 $. (\ref{68h}) applies for $ M_{R}(x) $ with $ x^{0}=0 $; however, since $D=e^{iH x^{0}}De^{-iH x^{0}}-Hx^{0}  $, $M_{R}(x^{i},x^{0})= e^{iH x^{0}} M_{R}(x^{i},0)e^{-iH x^{0}} $, from (\ref{68h}), 
\begin{equation}\label{68hf}
 -i[D,M_{R}(x)]=x^{\mu}\partial_{\mu}M_{R}(x)
\end{equation}
can be recovered.

$ M_{R}(x)  $ is a special kind of gauge transformation. In 3d Chern-Simons-matter theory, consider the generic local gauge transformation operator 
\begin{equation}
U(\alpha) =\exp \{-i \int d^{2}y \; tr [\alpha(y) \Lambda(y)]\}
\end{equation}
with $ \Lambda $ given by (\ref{Omega2}). At $  x^{0}=0$, $-i [D,  \Lambda]=2  \Lambda+ x^{i}\partial_{i} \Lambda  $. Under the scaling transformation $ e^{-i\lambda D} $, 
\begin{equation}
e^{-i\lambda D}\Lambda(x)e^{i\lambda D}=e^{2\lambda}\Lambda(e^{\lambda} x)\;,
\end{equation}
\begin{eqnarray}
\nonumber e^{-i\lambda D}U(\alpha)e^{i\lambda D} &=& \exp \{-i \int d^{2}y \; tr [e^{2\lambda}\alpha(y) \Lambda(e^{\lambda} y)]\}
=\exp \{-i \int d^{2} y \; tr [\alpha(e^{-\lambda}    y) \Lambda( y)]\} \\&=&U(\alpha')\;, 
\end{eqnarray}
where $ \alpha' $ is the gauge transformation parameter with $ \alpha'(y)=\alpha(e^{-\lambda}    y)  $. In $ M_{R}(x) $, $ \omega $ is scale invariant with $ \omega(e^{\lambda}x, y) = \omega(x, e^{-\lambda}    y) $, so
\begin{equation}
e^{-i\lambda D} M_{R}(x)  e^{i\lambda D} = M_{R}( e^{\lambda} x)  \;, 
\end{equation}
from which, (\ref{68h}) and then (\ref{68hf}) are obtained again.

As for the scaling transformation of $ \mathcal{M}_{R }(x)  $, from (\ref{328h}), 
\begin{equation}
e^{-i\lambda D} \mathcal{M}^{\alpha}_{R }(x) e^{i\lambda D}=\frac{1}{\sqrt{\mathcal{N}_{R}}}\int_{U \in \mathcal{G}} DU \; D^{\alpha }_{1 \;R }[u'(e^{\lambda}x)] U'M_{R/k}( e^{\lambda}x)U'^{-1}=\mathcal{M}^{\alpha}_{R }( e^{\lambda}x)\;,
\end{equation}
where for $ U $ with the transformation matrix $ u $, $ U' = e^{-i\lambda D} Ue^{i\lambda D}$ is the gauge transformation with the transformation matrix $ u' $, $ u'(x)=u(e^{-\lambda}   x) $, $ DU=DU' $. The classical conformal dimension of $ \mathcal{M}^{\alpha}_{R }(x) $, and similarly, of $ \mathcal{M}_{\alpha R }(x) $ and $ \mathcal{M}_{\hat{\alpha}(R_{1},\bar{R}_{2} )  }^{\alpha}(x)$, are all $ 0 $.

Note that for $\mathcal{M}_{R}  $ to have the conformal dimension $ 0 $ under the action of $ D $, we have selected a particular $ a_{i} $ in (\ref{55f}), although the generic $ a_{i} $ is given by (\ref{24f}). $ D $ is not gauge invariant. The gauge invariant version of $ D $ is $   \mathcal{D} $, which, when acting on physical operators, reduces to $ D $. $ \mathcal{M}_{R}  $'s constructed from the gauge equivalent $ a_{i} $ will all have the conformal dimension $ 0 $ under the action of $ \mathcal{D} $. We will return to this point in section \ref{so}.

\section{Supersymmetry transformation of monopole operators}\label{sus1}

In this section, we will study the supersymmetry transformation of monopole operators in 3d Super-Yang-Mills (SYM) theory and the ABJM theory. Concretely, we will calculate the action of the singular gauge transformation $ M_{R}   $ on supercharges, from which the commutator of the supercharges and $M_{R}  $ can be obtained.

The ordinary local gauge transformation operator $ U(\alpha) $ is characterized by the Lie algebra valued function $ \alpha $ which is well defined everywhere. With $ \alpha $ replaced by the multivalued function $ H_{\vec{m}}\omega $, we get $U(H_{\vec{m}}\omega)  = M_{R}   $. The gauge invariant operators are obviously invariant under the action of $ U(\alpha)  $. For example, for the supercharge $\mathcal{Q}  $, we have $U(\alpha) \mathcal{Q}U^{-1}(\alpha)  = \mathcal{Q}  $. However, not all gauge invariant operators would be invariant under the action of $ U(H_{\vec{m}}\omega)   $. In the following, we will compute the singular gauge transformation of the supercharges in SYM theory and ABJM theory explicitly.

Consider the 3d SYM theory with the gauge group $ U(N) $, where the fundamental fields are $(A_{i} , \Phi^{I} , \Psi)$ with the conjugate momentum $(\Pi^{i} , \Pi_{I} , \Pi )  $, $ i=1,2 $, $ I=1,\cdots,7 $.  Supercharges are gauge invariant operators composed by $ D_{i} \Phi^{I} $, $ [\Phi^{I} ,\Phi^{J} ] $, $ \Pi^{i} , \Pi_{I} , \Pi  , \Psi$, and $ F$, among which $ F $ transforms as 
\begin{equation}
M_{R}(x)F(y)M^{-1}_{R}(x)=e^{i H_{\vec{m}} \omega(x,y)}  F(y)e^{-i H_{\vec{m}} \omega(x,y) }+8 \pi H_{\vec{m}} \delta^{2}(x-y)
\end{equation}
under the action of $ M_{R}(x) $, while the other operators all transform as
\begin{equation}
M_{R}(x) D_{i} \Phi^{I}(y) M^{-1}_{R}(x)=e^{i H_{\vec{m}} \omega(x,y)}  D_{i} \Phi^{I} (y)e^{-i H_{\vec{m}} \omega(x,y) }\;. 
\end{equation}
As a result, for 
\begin{equation}
\mathcal{Q}=\int d^{2}y \; tr[(\frac{1}{4}F \epsilon_{ij}\Gamma^{ij}-\frac{i}{2}[\Phi_{I},\Phi_{J}]\Gamma^{IJ}+\Pi_{i}\Gamma^{0i}+\Pi_{I}\Gamma^{0I}+D_{i}\Phi_{J}\Gamma^{iJ})\Gamma^{0}\Psi
]\;,
\end{equation}
where $ \Gamma $'s are $ 32 \times 32 $ gamma matrices in 10d spacetime, we have 
\begin{equation}
M_{R}(x)\mathcal{Q}M^{-1}_{R}(x)=\mathcal{Q} +2 \pi \epsilon_{ij}\Gamma^{ij}\Gamma^{0}tr[H_{\vec{m}} \Psi(x)]
\end{equation}
and 
\begin{equation}
[\mathcal{Q}, M_{R}(x)]=-2 \pi \epsilon_{ij}\Gamma^{ij}\Gamma^{0}tr[H_{\vec{m}}  \Psi(x)]M_{R}(x)\;.
\end{equation}

In fact, for Yang-Mills-matter theory, $M_{R}(x)  $ given by (\ref{1m}) takes the same form as the ordinary local gauge transformation operator with $ a_{i} =\partial_{i}\omega$ satisfied as well. The obstruction for $M_{R}(x)    $ to be a genuine gauge transformation is $ \epsilon^{ij} \partial_{i}a_{j} (x,y)= 4 \pi \delta^{2}(x-y) \neq 0$, so when computing the supersymmetry variation of $M_{R}(x)  $, we only need to consider terms producing $ \epsilon^{ij} \partial_{i}a_{j} $. The supersymmetry variation of $\Pi^{i}  $ is 
\begin{equation}\label{66a1}
[\mathcal{Q},\Pi^{i}]=-i [\Gamma^{ji}\Gamma^{0}\partial_{j}\Psi+ \cdots]\;,
\end{equation}
where ``$ \cdots $'' are terms with no derivatives, so
\begin{eqnarray}
\nonumber [\mathcal{Q},M_{R}(x)] &=&\int d^{2}y \; tr H_{\vec{m}}(\Gamma^{ji}\Gamma^{0}\partial_{j}\Psi a_{i} )M_{R}(x)   =\int d^{2}y \; tr H_{\vec{m}}(-\Gamma^{ji}\Gamma^{0}\Psi \partial_{j}a_{i} )M_{R}(x)  \\\nonumber  &=&-2\pi \epsilon_{ij}\Gamma^{ij}\Gamma^{0}tr [H_{\vec{m}}\Psi(x) ]M_{R}(x)
\end{eqnarray}
is obtained again.

It is possible to construct the scalar-dressed monopole operator preserving part of the supersymmetry. Since 
\begin{equation}
[\mathcal{Q},\Phi^{I}]=-i \Gamma^{I}\Psi\;, 
\end{equation}
for the operator $ \exp \{-4\pi tr[H_{\vec{m}}\Phi^{I}(x) ]\} M_{R}(x)$, the supersymmetry variation is 
\begin{eqnarray}\label{789}
\nonumber && [\bar{\epsilon}\mathcal{Q},\exp \{-4\pi tr [H_{\vec{m}}\Phi^{I}(x) ]\}M_{R}(x)]  \\&=&-4\pi \bar{\epsilon}(-i  \Gamma^{I}+\Gamma^{1}\Gamma^{2}\Gamma^{0})tr [H_{\vec{m}}\Psi(x) ]\exp \{-4\pi tr [H_{\vec{m}}\Phi^{I}(x) ]\}M_{R}(x)\;.
\end{eqnarray}
(\ref{789})$ =0 $ for $ \epsilon $ satisfying $\bar{\epsilon}\Gamma^{1}\Gamma^{2}\Gamma^{0} \Gamma^{I}=i\bar{\epsilon}  $, so $ \exp \{-4\pi tr [H_{\vec{m}}\Phi^{I}(x) ]\}M_{R}(x) $ is a $ 1/2 $ BPS operator.

Next, consider the $ U(N)_{k} \times U(N)_{-k} $ ABJM theory with $ \mathcal{N} =6$ supersymmetry. Following the convention in \cite{tw10b}, the field content consists of complex scalars $ X^{a}_{A \hat{a}} $, spinors $\Psi^{Aa}_{ \hat{a}}  $, and their adjoints $ X_{a}^{A \hat{a}}  $, $ \Psi_{Aa}^{ \hat{a}}   $, transforming as $ (N,\bar{N})  $ and $(\bar{N},N)  $ representations of the gauge group. $ A=1,\cdots,4 $, $ a,\hat{a} =1,\cdots,N$. The $ U(N) $ gauge fields are Hermitian matrices $ A^{a}_{b} $ and $ \hat{A}_{\hat{b}}^{\hat{a}} $. The covariant derivatives are
\begin{equation}
D_{\mu}X_{A}=\partial_{\mu}X_{A}-i(A_{\mu}X_{A}-X_{A}\hat{A}_{\mu})\;,\;\;\;\;\;\;\;\;D_{\mu}X^{A}=\partial_{\mu}X^{A}-i(\hat{A}_{\mu}X^{A}-X^{A}A_{\mu})
\end{equation}
with similar formulas for the spinors. $ \mu=0,1,2 $. The action of the ABJM theory is \cite{tw10b} 
\begin{equation}\label{69i}
S=S_{kin}+S_{SC}+S_{F}+S_{S}
\end{equation}
with the kinetic term
\begin{equation}
S_{kin}=\frac{k}{2\pi}\int d^{3}x\;tr(-D^{\mu}X^{A}D_{\mu}X_{A}+i \bar{\Psi}_{A}\gamma^{\mu}D_{\mu}\Psi^{A} )\;,
\end{equation}
the Chern-Simons term
\begin{equation}
S_{SC}=\frac{k}{2\pi}\int d^{3}x\;\epsilon^{\mu\nu\lambda} tr(\frac{1}{2}A_{\mu}\partial_{\nu}A_{\lambda}-\frac{i}{3}A_{\mu}A_{\nu}A_{\lambda}-\frac{1}{2}\hat{A}_{\mu}\partial_{\nu}\hat{A}_{\lambda}+\frac{i}{3}\hat{A}_{\mu}\hat{A}_{\nu}\hat{A}_{\lambda})\;,
\end{equation}
the fermionic interaction 
\begin{eqnarray}\label{712z}
\nonumber S_{F} &=&i \int d^{3}x\;  \epsilon^{ABCD} tr(\bar{\Psi}_{A}X_{B}\Psi_{C}X_{D}) - \epsilon_{ABCD} tr(\bar{\Psi}^{A}X^{B}\Psi^{C}X^{D})    \\  &+&   tr( \bar{\Psi}^{A} \Psi_{A} X_{B}X^{B}-  \bar{\Psi}_{A} \Psi^{A}X^{B} X_{B} +2 \bar{\Psi}_{A} \Psi^{B}X^{A} X_{B} -2\bar{\Psi}^{B} \Psi_{A} X_{B}X^{A} )\;,
\end{eqnarray}
and the scalar potential 
\begin{eqnarray}\label{613i}
\nonumber S_{S} &=&\frac{1}{3} \int d^{3}x\;  tr(X^{A}X_{A}X^{B}X_{B}X^{C}X_{C}+X_{A}X^{A}X_{B}X^{B}X_{C}X^{C}
 \\  &+&  4 X_{A}X^{B}X_{C}X^{A}X_{B}X^{C}-6X^{A}X_{B}X^{B}X_{A}X^{C}X_{C})\;.
\end{eqnarray}
$ \gamma^{\mu} $'s are $ 2 \times 2 $ Dirac matrices satisfying $ \{\gamma^{\mu},\gamma^{\nu}\}=2 \eta^{\mu\nu}$. $ \bar{\Psi}^{A} $ and $ \bar{\Psi}_{A} $ are transpositions of $ \Psi^{A} $ and $\Psi_{A} $ multiplied by $ \gamma^{0} $.

Under the $ \mathcal{N} =6$ supersymmetry transformation \cite{tw10b},
\begin{eqnarray}
\label{817a}   &&  \delta X_{A}=i \Gamma^{I}_{AB}\bar{\Psi}^{B}\varepsilon^{I} \;,  \\  &&   \delta \Psi_{A}=\Gamma^{I}_{AB}\gamma^{\mu}\varepsilon^{I}D_{\mu}X^{B}+N^{I }_{A}\varepsilon^{I}\;,\\\label{817} &&  \delta A_{\mu }=-\Gamma^{I}_{AB}\bar{\Psi}^{A}
\gamma_{\mu}  \varepsilon^{I} X^{B}-\tilde{\Gamma}^{IAB}X_{B} \bar{\Psi}_{A} \gamma_{\mu}\varepsilon^{I}\;,\\\label{817b}  &&  \delta \hat{A}_{\mu}=-\Gamma^{I}_{AB}X^{B}\bar{\Psi}^{A}
\gamma_{\mu}  \varepsilon^{I}-\tilde{\Gamma}^{IAB} \bar{\Psi}_{A} \gamma_{\mu}\varepsilon^{I}X_{B}\;,
\end{eqnarray}
where 
\begin{equation}
N^{I}_{A}=\Gamma^{I}_{AB}(X^{C}X_{C}X^{B}-X^{B}X_{C}X^{C})-2\Gamma^{I}_{BC}X^{B}X_{A}X^{C}\;.
\end{equation}
$ \Gamma^{I} $ with $ I=1,\cdots,6 $ are $ 4 \times 4 $ matrices satisfying $\Gamma^{I}_{AB}=-\Gamma_{BA}^{I}  $, 
\begin{equation}
\tilde{\Gamma}^{IAB}=\frac{1}{2}\epsilon^{ABCD}\Gamma_{CD}^{I}=-(\Gamma_{AB}^{I})^{*}\;,
\end{equation}
\begin{equation}
\Gamma^{I}\tilde{\Gamma}^{J}+\Gamma^{J}\tilde{\Gamma}^{I}=2\delta^{IJ}\;.
\end{equation}
The $ 6 $ supercharges are given by \cite{tw10b} 
\begin{equation}\label{621z}
\mathcal{Q}^{I}=-\frac{k}{2\pi}\int d^{2}x \; tr[(-\Gamma^{I}_{AB}\gamma^{\nu}D_{\nu}X^{B} +N^{I}_{A} )\gamma_{0}\Psi^{A}]+tr[(\tilde{\Gamma}^{IAB}\gamma^{\nu}D_{\nu}X_{B}+N^{IA})\gamma_{0}\Psi_{A}]\;,
\end{equation}
$ N^{IA}=(N^{I}_{A})^{+} $. The $ 6 $ conformal supercharges are given by \cite{tw10b} 
\begin{eqnarray}
\nonumber\mathcal{S}^{I} &=&-\frac{k}{2\pi}\int d^{2}x \; tr[\gamma^{\mu}x_{\mu}(-\Gamma^{I}_{AB}\gamma^{\nu}D_{\nu}X^{B} +N^{I}_{A} )\gamma_{0}\Psi^{A}]+tr[\gamma^{\mu}x_{\mu}(\tilde{\Gamma}^{IAB}\gamma^{\nu}D_{\nu}X_{B}+N^{IA})\gamma_{0}\Psi_{A}]
 \\  &-&  \Gamma^{I}_{AB} tr(X^{B}\gamma_{0}\Psi^{A})+\tilde{\Gamma}^{IAB} tr(X_{B}\gamma_{0}\Psi_{A})
 \;.
\end{eqnarray}

In parallel with the discussion for 3d SYM theory, $ \mathcal{Q}^{I} $ and $  \mathcal{S}^{I} $ are gauge invariant and do not contain $ F $, and hence are invariant under the action of $ M_{R}(x) $.
\begin{equation}
M_{R}(x)  \mathcal{Q}^{I} M^{-1}_{R}(x)= \mathcal{Q}^{I}  \;,\;\;\;\;\; [\mathcal{Q}^{I}, M_{R}(x)]=0  \;,\;\;\;\;\;M_{R}(x)  \mathcal{S}^{I} M^{-1}_{R}(x)= \mathcal{S}^{I}  \;,\;\;\;\;\; [\mathcal{S}^{I}, M_{R}(x)]=0  \;.
\end{equation}
$M_{R}(x)    $ is SUSY invariant.

In ABJM theory, $ M_{R}(x) $ is given by (\ref{UU}), in which terms involving $ a_{i} $ are $ \epsilon^{ij} A_{j}a_{i}$ and $ \epsilon^{ij} \hat{A}_{j}a_{i}  $, so to compute the supersymmetry variation of $ M_{R}(x)  $, we only need to consider $  \delta A$ and $ \delta  \hat{A}$. However, in (\ref{817}) and (\ref{817b}), $ \delta A $ and $  \delta  \hat{A} $ do not contain derivatives, so in $ [\mathcal{Q}^{I}, M_{R}(x)] $ and $  [\mathcal{S}^{I}, M_{R}(x)]  $, $ \epsilon^{ij} \partial_{i}a_{j} $ cannot be produced. The difference between $M_{R}(x)  $ and $ U(\alpha) $ does not have the manifestation here, so just as $U(\alpha)   $, $ M_{R}(x)   $ commutes with the supercharge.

Finally, for $ \mathcal{M}_{\hat{\alpha}(R_{1},\bar{R}_{2} )  }^{\alpha}(x)   $ in representation $ (R_{1},  \bar{R}_{2} ) $ constructed in (\ref{435}), $ [\mathcal{Q}^{I}, M_{[R_{1},\bar{R}_{2}]}(x)]=0 $, $ [\mathcal{S}^{I}, M_{[R_{1},\bar{R}_{2}]}(x)]=0 $, $ [\mathcal{Q}^{I}, U] =[\mathcal{Q}^{I}, \hat{U}]=0$, $ [\mathcal{S}^{I}, U] =[\mathcal{S}^{I}, \hat{U}]=0$, so
\begin{equation}
[\mathcal{Q}^{I},  \mathcal{M}_{\hat{\alpha}(R_{1},\bar{R}_{2} )  }^{\alpha}(x)   ]=0 \;,\;\;\;\;\;\;\; [\mathcal{S}^{I},  \mathcal{M}_{\hat{\alpha}(R_{1},\bar{R}_{2} )  }^{\alpha}(x)    ]=0  \;.
\end{equation}

\section{Derivative and covariant derivative of monopole operators}\label{der}

In this section, we will compute the derivative and covariant derivative of the monopole operators $ M_{R} (x)$ and $ \mathcal{M}_{R} (x) $ in 3d Yang-Mills-matter theory and Chern-Simons-matter theory. We find that in Chern-Simons-matter theory, $ M_{R} (x)$ and $ \mathcal{M}_{R} (x) $ are covariantly constant.

The monopole operator $ M_{R}(x) $ at the point $ x $ is characterized by the function $H_{\vec{m}} a_{i}(x,y) $. Under the infinitesimal translation $ x\rightarrow x+\xi $, $M_{R}(x) \rightarrow M_{R}(x+\xi )   $, $ a_{i}(x,y) \rightarrow a_{i}(x+\xi ,y)=  a_{i}(x ,y-\xi)$.

In 3d Yang-Mills-matter theory, $ M_{R}(x) $ is given by 
\begin{equation}
M_{R}(x)=\exp \{i \int d^{2}y \; tr H_{\vec{m}}[\Pi^{i}(y)a_{i}(x,y)+i [A_{i}(y),\Pi^{i}(y)]\omega(x,y)  +\rho(y)\omega(x,y)  ]\}\;.
\end{equation}
From (\ref{38u}), the derivative of $ M_{R}(x) $ is
\begin{eqnarray}\label{84e1}
\nonumber \partial_{i}M_{R}(x) &= &
-i \int d^{2}y \; tr H_{\vec{m}}[\Pi^{j} \partial_{i} a_{j}+i [A_{j},\Pi^{j}] \partial_{i} \omega  +\rho\partial_{i}  \omega  ]
M_{R}(x) 
\\  &= &\{4 \pi  i  \epsilon_{ji} tr [H_{\vec{m}} \Pi^{j}(x)]
+i \int d^{2}y \; tr[ H_{\vec{m}}a_{i}(x,y)\Lambda(y)] \}
M_{R}(x) \;,
\end{eqnarray}
where $\Lambda =  \partial_{i}  \Pi^{i}      -i [A_{i},\Pi^{i}] -\rho$. We can also calculate the covariant derivative of $ M_{R}(x) $. Consider the local gauge transformation operator $ U(\alpha) =\exp \{-iG( \alpha)\}$ with the generator
\begin{equation}
G( \alpha)=\int d^{2}y \; tr[ \alpha(y)\Lambda(y)] \;.
\end{equation}
Under the action of $ M_{R}(x)  $, $ G( A_{i}) $ transforms as 
\begin{equation}
M_{R}(x) G( A_{i})M^{-1}_{R}(x)  =G( A_{i})+\int d^{2}y \; tr[  H_{\vec{m}} a_{i}(x,y)\Lambda(y) ]=G( A_{i})+G(  H_{\vec{m}}  a_{i})\;,
\end{equation}
so
\begin{equation}\label{gm}
[G( A_{i}),M_{R}(x)  ]=-G(  H_{\vec{m}}  a_{i}) M_{R}(x)  \;.
\end{equation}
The covariant derivative of $ M_{R}(x) $ is 
\begin{equation}\label{76e}
 D_{i}M_{R}(x)= \partial_{i}M_{R}(x) +i [G( A_{i}),M_{R}(x)  ]=4 \pi  i  \epsilon_{ji} tr [H_{\vec{m}} \Pi^{j} (x)]
M_{R}(x)\;.
\end{equation}
Compared with (\ref{84e1}), the right-hand side of (\ref{76e}) only contains local operators.

Of course, the derivative could be directly obtained from the action of the momentum operator $ P_{i} $. For example, in a 3d $ U(N)  $ Yang-Mills-matter theory with the matter part composed of a scalar $ \phi $ and a spinor $ \psi $, both in adjoint representation of $ U(N) $, 
\begin{equation}
P_{i}=\int d^{2}y\; tr(\Pi_{\phi}\partial_{i}\phi+\Pi_{\psi} \partial_{i}\psi+\Pi^{j}\partial_{i}A_{j})\;,
\end{equation}
where $ \Pi_{\phi} $ and $\Pi_{\psi}  $ are conjugate momenta of $\phi $ and $\psi $. For $ f =(\Pi_{\phi}, \phi, \Pi_{\psi} , \psi, \Pi^{i}, A_{i})$, $  \partial_{i}f(x)=i[P_{i},f(x) ] $. Under the action of $ M_{R }(x)  $,
\begin{equation}
M_{R }(x) P_{i} M^{-1}_{R }(x)= P_{i}-G(H_{\vec{m}}a_{i})-4 \pi \epsilon_{ji}tr[H_{\vec{m}}\Pi^{j}(x)]\;,
\end{equation}
so
\begin{equation}\label{pm}
 \partial_{i}M_{R }(x)=i[P_{i},M_{R }(x) ]=  \{4 \pi i \epsilon_{ji} tr[H_{\vec{m}}\Pi^{j}(x)]+iG(  H_{\vec{m}}  a_{i})\}M_{R }(x)\;,
\end{equation}
which is consistent with (\ref{84e1}).

The gauge invariant completion of $P_{i}  $ is 
\begin{equation}\label{710}
\mathcal{P}_{i}=\int d^{2}y\; tr(\Pi_{\phi}D_{i}\phi+\Pi_{\psi} D_{i}\psi+
\Pi^{j}F_{ij})
=P_{i}+\int d^{2}y\; tr(A_{i}\Lambda)=P_{i}+G(A_{i})\;. 
\end{equation}
When $ f =(\Pi_{\phi}, \phi, \Pi_{\psi} , \psi) $, $ [ \mathcal{P}_{i},f(x)]=-i D_{i}f(x) $. For the gauge field $ A_{j} $, 
\begin{equation}\label{710d}
[ \mathcal{P}_{i},A_{j}(x)]=-i F_{ij}(x)\;.
\end{equation}
From (\ref{gm}), (\ref{pm}) and (\ref{710}), (\ref{76e}) is recovered.

In 3d Chern-Simons-matter theory with the level $ k $, the monopole operator $ M_{R}(x) $ is 
\begin{equation}\label{75}
M_{R}(x) =\exp \{\frac{ik}{2\pi} \int d^{2}y \; tr H_{\vec{m}}[ \epsilon^{ij}A_{j}(y)a_{i}(x,y)+\frac{i}{2}\epsilon^{ij}[A_{i}(y),A_{j}(y)]\omega(x,y)+\frac{2\pi}{k}\rho (y)\omega(x,y) ]\}\;.
\end{equation}
Still, consider a Chern-Simons-matter theory with the gauge group $ U(N) $ and the field content $ (A_{i},\phi,\psi) $. The momentum operator is 
\begin{equation}
P_{i}=\int d^{2}y\; tr(\Pi_{\phi}\partial_{i}\phi+\Pi_{\psi}\partial_{i}\psi-\frac{k}{4\pi}A_{i}F)\;.
\end{equation}
For $ f =(\Pi_{\phi},\phi,\Pi_{\psi},\psi,A_{i})$, $  \partial_{i}f(x)=i[P_{i},f(x) ] $. Under the action of $M_{R }(x)   $,
\begin{equation}
M_{R }(x) P_{i} M^{-1}_{R }(x)= P_{i}-\int d^{2}y\; tr[ H_{\vec{m}} a_{i} (x,y) \Lambda(y)] -2k tr[H_{\vec{m}}A_{i} (x)+H_{\vec{m}}H_{\vec{m}}a_{i} (x,x)]\;,
\end{equation}
where
\begin{equation}
\Lambda=\frac{k}{4\pi}F-i([\phi,\Pi_{\phi}]-\{\psi,\Pi_{\psi}\})\;,
\end{equation}
so
\begin{equation}\label{815}
 \partial_{i}M_{R }(x) = i[P_{i},M_{R }(x) ]= iG(H_{\vec{m}} a_{i} )M_{R }(x) +2ki tr[H_{\vec{m}}A_{i} (x)+H_{\vec{m}}H_{\vec{m}}a_{i} (x,x)]M_{R }(x) \;.
\end{equation}

The gauge invariant momentum operator is
\begin{equation}\label{7190}
\mathcal{P}_{i}=\int d^{2}y\; tr(\Pi_{\phi}D_{i}\phi+\Pi_{\psi}D_{i}\psi)
=P_{i}+\int d^{2}y\; tr(A_{i}\Lambda)=P_{i}+G(A_{i})\;.
\end{equation}
In particular, in pure CS theory, $\mathcal{P}_{i}=0  $, and moreover, with $ A_{0} =0$, $ H=0 $ \cite{CSS}. When $ f =(\Pi_{\phi},\phi,\Pi_{\psi},\psi) $, $ [ \mathcal{P}_{i},f(x)]=-i D_{i}f(x) $. For the gauge field $ A_{j} $, 
 \begin{equation}
 [\mathcal{P}_{i},A_{j}]=\frac{2\pi }{k}\epsilon_{ij}([\phi,\Pi_{\phi}]-\{\psi,\Pi_{\psi}\})
 \;.
\end{equation}
in contrast to (\ref{710d}). $ \mathcal{P}_{i}  $ is gauge invariant and does not contain $ F $, so
\begin{equation}
M_{R }(x) \mathcal{P}_{i} M^{-1}_{R }(x)= \mathcal{P}_{i}\;,\;\;\;\;\;\;\;
D_{i}M_{R }(x) = i[\mathcal{P}_{i} ,M_{R }(x) ]= 0\;,
\end{equation}
$ M_{R }(x)  $ is covariantly constant. On the other hand, in Yang-Mills-matter theory, $ M_{R }(x)  $ is not covariantly constant merely because $\mathcal{P}_{i}  $ given by (\ref{710}) has $ F $ involved.

Next, let us compute the derivative and the covariant derivative of the monopole operator $ \mathcal{M}_{R }(x) $ in representation $ R $. 
\begin{equation}
\mathcal{M}^{\alpha}_{R }(x) =\frac{1}{\sqrt{N_{R}}}\int DU \; D^{\alpha }_{1 \;R }[u(x)]  UM_{R/k}(x)U^{-1}\;. 
\end{equation}
Aside from $ M_{R/k}(x) $, $ D^{\alpha }_{1 \;R }[u(x)]   $ also has the explicit $ x $-dependence, so we should check whether $\partial_{i} \mathcal{M}_{R }(x) $ can still be realized by the action of $ P_{i} $.

The action of the translation operator $\exp \{i P_{i}\xi^{i}\}  $ transforms $ M_{R}(x) $ into
\begin{equation}
M_{R}(x+\xi)=\exp \{i P_{i}\xi^{i}\} M_{R}(x) \exp \{-i P_{i}\xi^{i}\}\;.
\end{equation}
For $ \mathcal{M}^{\alpha}_{R }(x) $, we have
\begin{equation}
\exp \{i P_{i}\xi^{i}\} \mathcal{M}^{\alpha}_{R }(x)\exp \{-i P_{i}\xi^{i}\}=\frac{1}{\sqrt{N_{R}}}\int DU' \; D^{\alpha }_{1 \;R }[u'(x+\xi)]  U'    M_{R/k}(x+\xi)   U'^{-1}=
\mathcal{M}^{\alpha}_{R }(x+\xi)\;,
\end{equation}
where 
\begin{equation}
U'=\exp \{i P_{i}\xi^{i}\} U  \exp \{-i P_{i}\xi^{i}\}
\end{equation}
is the gauge transformation with the related transformation matrix $ u'(x)=u(x-\xi) $, $ DU=DU' $. So
\begin{equation}
 \partial_{i}\mathcal{M}^{\alpha}_{R }(x) = i[P_{i},\mathcal{M}^{\alpha}_{R }(x) ]
\end{equation}
is still valid.

The covariant derivative operator $ \mathcal{P}_{i}  $ is gauge invariant, $[U,\mathcal{P}_{i}  ]=0  $,
\begin{equation}\label{730}
D_{i}\mathcal{M}^{\alpha}_{R }(x) = i[ \mathcal{P}_{i},\mathcal{M}^{\alpha}_{R }(x)] =\frac{i}{\sqrt{N_{R}}}\int DU \; D^{\alpha }_{1 \;R }[u(x)]  U[ \mathcal{P}_{i},M_{R/k}(x)]   U^{-1}=0\;. 
\end{equation}
$ \mathcal{M}^{\alpha}_{R }(x)  $ is also covariantly constant.

Since $ \mathcal{M}^{\alpha}_{R }(x)  $ is an operator in representation $ R $, $ D_{i}\mathcal{M}^{\alpha}_{R }(x) $ can be written explicitly. Consider the gauge invariant operator $O_{\alpha R}(x)\mathcal{M}^{\alpha}_{R }(x)   $ with $ O_{\alpha R}(x) $ in representation $ \bar{R} $ constructed from the matter fields. 
\begin{equation}
D_{i} O_{\alpha R}(x) = i[ \mathcal{P}_{i}, O_{\alpha R}(x) ] =\partial_{i}O_{\alpha R}(x) + i  O_{\beta R}(x)A^{\beta}_{\alpha\; i}(x)\;.
\end{equation}
When acting on the gauge invariant operators, $\mathcal{P}_{i}  $ reduces to $ P_{i} $, so 
\begin{eqnarray}
\nonumber \partial_{i} [O_{\alpha R}(x)\mathcal{M}^{\alpha}_{R }(x)] &=&i[ \mathcal{P}_{i},  O_{\alpha R}(x)]  \mathcal{M}^{\alpha}_{R }(x)+i O_{\alpha R}(x) [ \mathcal{P}_{i},  \mathcal{M}^{\alpha}_{R }(x)  ]  \\ &=&[\partial_{i}O_{\alpha R}(x) + i  O_{\beta R}(x)A^{\beta}_{\alpha\; i}(x)]\mathcal{M}^{\alpha}_{R }(x)+i O_{\alpha R}(x) [ \mathcal{P}_{i},  \mathcal{M}^{\alpha}_{R }(x)  ]   \;.
\end{eqnarray}
The covariant derivative of $ \mathcal{M}^{\alpha}_{R }(x) $ is 
\begin{equation}
D_{i}\mathcal{M}^{\alpha}_{R }(x) = i[ \mathcal{P}_{i},\mathcal{M}^{\alpha}_{R }(x)]=\partial_{i}\mathcal{M}^{\alpha}_{R }(x) - i  A^{\alpha}_{\beta\; i}(x) \mathcal{M}^{\beta}_{R }(x)=0\;.
\end{equation}

For the 3d Chern-Simons-matter theory with the gauge group $ U(N)_{k} \times U(N)_{-k} $, $ \mathcal{M}_{\hat{\alpha}R  }^{\alpha}(x) $ in representation $ (R,\bar{R}) $ is covariantly constant with 
\begin{equation}
D_{i}\mathcal{M}_{\hat{\alpha}R  }^{\alpha}(x)  =\partial_{i}\mathcal{M}^{\alpha}_{\hat{\alpha}R }(x) - i  A^{\alpha}_{\beta\; i}(x) \mathcal{M}^{\beta}_{\hat{\alpha}R }(x)+ i   \mathcal{M}^{\alpha}_{\hat{\beta}R }(x)\hat{A}^{\hat{\beta}}_{\hat{\alpha}\; i}(x)=0\;.
\end{equation}

We should also consider the derivative and the covariant derivative along the time direction. Since $ A_{0} =\hat{A}_{0}=0$, $ D_{0} =\partial_{0}$, 
\begin{equation}
D_{0} M_{R}(x)=\partial_{0}M_{R}(x)=i[H, M_{R}(x)]\;,
\end{equation}
where $ H $ is the Hamiltonian. In Chern-Simons-matter theory, $ H $ is gauge invariant and does not contain $ F $, and thus would commute with $ M_{R}(x) $. $D_{0} M_{R}(x)=\partial_{0}M_{R}(x)=0  $. $ [U,H] =[\hat{U},H]=0$, so $D_{0} \mathcal{M}^{\alpha}_{R }(x)=\partial_{0}\mathcal{M}^{\alpha}_{R }(x)=0  $, $ D_{0}\mathcal{M}_{\hat{\alpha}R  }^{\alpha}(x)  =\partial_{0}\mathcal{M}_{\hat{\alpha}R  }^{\alpha}(x)  =0   $.

Finally, in the above discussion, we take the particular Yang-Mills-matter theory and Chern-Simons-matter theory as the examples, but the result is independent of the matter content. We may conclude that in the generic Chern-Simons-matter theories, the monopole operators are covariantly constant.

\section{The conservation of the $SO(8)  $ R-symmetry current  in ABJM theory}\label{so}

ABJM theory has the manifest $ SU(4) \times U(1)_{J}$ global symmetry with $ SU(4) $ and $ U(1)_{J} $ realized by $ 15 $ traceless currents \cite{CUR}
\begin{eqnarray}\label{jab1}
\nonumber j^{A}_{\mu B} &=&-\frac{ik}{2\pi}\{[X_{a}^{A \hat{a}}(D_{\mu}X_{B})^{a}_{\hat{a}}-(D_{\mu}X^{A})_{a}^{\hat{a}}X_{B\hat{a}}^{a}+i \bar{\Psi}^{Aa}_{\hat{a}}\gamma_{\mu} \Psi_{Ba}^{ \hat{a}}] \\ &-&\frac{1}{4}\delta_{B}^{A}[X_{a}^{C \hat{a}}(D_{\mu}X_{C})^{a}_{\hat{a}}-(D_{\mu}X^{C})_{a}^{\hat{a}}X_{C\hat{a}}^{a}+i \bar{\Psi}^{Ca}_{\hat{a}}\gamma_{\mu} \Psi_{Ca}^{ \hat{a}}]\}
\end{eqnarray}
and a single current 
\begin{equation}\label{92l}
j_{\mu}=\frac{1}{4\pi}\epsilon_{\mu\nu\lambda}( tr F^{\nu\lambda}+tr \hat{F}^{\nu\lambda})
\end{equation}
whose conservation follows from the equations of motion and the Bianchi identity. Using the Gauss constraint, (\ref{92l}) can also be written as the global $ U(1) $ charge current.

When $ k=1,2 $, with the help of the monopole operator, $ 12 $ extra currents \cite{S2, CUR} 
\begin{equation}\label{jab}
j^{AB}_{\mu }=-\frac{ik}{4\pi}\mathcal{M}_{\hat{a} \hat{b} }^{ab}[X_{a}^{A \hat{a}}(D_{\mu}X^{B})_{b}^{\hat{b}}-(D_{\mu}X^{A})_{a}^{\hat{a}}X_{b}^{B \hat{b}}+\frac{i}{2}\epsilon^{ABCD} \bar{\Psi}_{Ca}^{\hat{a}}\gamma_{\mu} \Psi_{Db}^{ \hat{b}}]
\end{equation}
and their adjoints $j_{\mu AB} =j^{AB+}_{\mu } $ can be constructed, which, if are conserved, will offer the off-diagonal charges, making the original $ SU(4) \times U(1)_{J}$ symmetry enhanced to $ SO(8) $.

Let us consider the current conservation equation for $ j^{AB}_{\mu } $. In Appendix \ref{AAA}, with the equations of motion plugged in, $ \partial^{\mu}j^{AB}_{\mu }  $ could be written as 
\begin{eqnarray}
\nonumber &&\partial^{\mu}j^{AB}_{\mu } \\ \nonumber &=&-\frac{ik}{4\pi}(D^{\mu}\mathcal{M})_{\hat{a} \hat{b} }^{ab}[X_{a}^{A \hat{a}}(D_{\mu}X^{B})_{b}^{\hat{b}}-(D_{\mu}X^{A})_{a}^{\hat{a}}X_{b}^{B \hat{b}}+\frac{i}{2}\epsilon^{ABCD} \bar{\Psi}_{Ca}^{\hat{a}}\gamma_{\mu} \Psi_{Db}^{ \hat{b}}]  \\ \nonumber &-&(W^{D}_{C} \mathcal{M} )_{\hat{a} \hat{b} }^{ab}[X_{a}^{B \hat{a}}(W^{C}_{D} X^{A})_{b}^{\hat{b}}-X_{a}^{A \hat{a}} (W^{C}_{D} X^{B})_{b}^{\hat{b}}]-\frac{1}{2}(W^{C}_{C} \mathcal{M} )_{\hat{a} \hat{b} }^{ab}[X_{a}^{A \hat{a}} (W^{D}_{D} X^{B})_{b}^{\hat{b}}-X_{a}^{B \hat{a}} (W^{D}_{D} X^{A})_{b}^{\hat{b}}]\\ \nonumber &-&  \frac{i}{2}\bar{\Psi}^{ \hat{b}}_{Cb}(V_{DE} \mathcal{M})_{\hat{a} \hat{b} }^{ab}(\epsilon^{ABCD}X_{a}^{E \hat{a}} -2 \epsilon^{ABCE}X_{a}^{D \hat{a}})
\\ \nonumber  &-& 
\frac{i}{2}[(U^{BC}  \mathcal{M})_{\hat{a} \hat{b} }^{ab}  X_{a}^{A \hat{a}}  -( U^{AC}  \mathcal{M} )_{\hat{a} \hat{b} }^{ab}X_{a}^{B \hat{a}}  +2(U^{CA}  \mathcal{M} )_{\hat{a} \hat{b} }^{ab}X_{a}^{B \hat{a}}   -2(U^{CB}   \mathcal{M} )_{\hat{a} \hat{b} }^{ab}X_{a}^{A \hat{a}}]\Psi^{\hat{b}}_{C b} 
\;.\\
\end{eqnarray}
$ D^{\mu}\mathcal{M} $ is the covariant derivative of $ \mathcal{M} $:
\begin{equation}
(D^{\mu}\mathcal{M})_{\hat{a} \hat{b} }^{ab}=\partial^{\mu}\mathcal{M}_{\hat{a} \hat{b} }^{ab}-iA^{\mu a}_{ c}\mathcal{M}_{\hat{a} \hat{b} }^{cb}-iA^{\mu b}_{c}\mathcal{M}_{\hat{a} \hat{b} }^{ac}+i\mathcal{M}_{\hat{c} \hat{b} }^{ab}\hat{A}^{\mu \hat{c} }_{ \hat{a} }+i\mathcal{M}_{\hat{a} \hat{c} }^{ab}\hat{A}^{\mu\hat{c} }_{\hat{b} }\;.
\end{equation}
$ W^{C}_{D} $ is a $ U(N)\times U(N) $ gauge variation whose actions on $ X^{ B}   $ and $  \mathcal{M} $ are given by 
\begin{equation}
(W^{C}_{D} X^{B})_{a}^{ \hat{a}}\equiv (X^{C}X_{D})^{\hat{a}}_{\hat{b}}X_{a}^{B\hat{b}}-X^{B\hat{a}}_{b}(X_{D}X^{C})_{a}^{b}
\end{equation}
and 
\begin{equation}
(W^{C}_{D} \mathcal{M}  )_{\hat{a} \hat{b} }^{ab} \equiv (X_{D}X^{C})_{c}^{a}   \mathcal{M}_{\hat{a} \hat{b} }^{cb}+(X_{D}X^{C})_{c}^{b}   \mathcal{M}_{\hat{a} \hat{b} }^{ac}- \mathcal{M}_{\hat{c} \hat{b} }^{ab} (X^{C}X_{D})^{\hat{c}}_{\hat{a}}- \mathcal{M}_{\hat{a} \hat{c} }^{ab} (X^{C}X_{D})^{\hat{c}}_{\hat{b}}\;,
\end{equation}
respectively. 
$ V_{CD} $ and $  U^{CB}  $ are $ U(N)\times U(N) $ gauge variations whose actions on $ \mathcal{M}  $ are given by 
\begin{equation}
(V_{CD} \mathcal{M}  )_{\hat{a} \hat{b} }^{ab} \equiv (X_{D} \Psi_{C})_{c}^{a}   \mathcal{M}_{\hat{a} \hat{b} }^{cb}+(X_{D} \Psi_{C})_{c}^{b}   \mathcal{M}_{\hat{a} \hat{b} }^{ac}- \mathcal{M}_{\hat{c} \hat{b} }^{ab} (\Psi_{C} X_{D})^{\hat{c}}_{\hat{a}}- \mathcal{M}_{\hat{a} \hat{c} }^{ab} (\Psi_{C} X_{D})^{\hat{c}}_{\hat{b}}
\end{equation}
and
\begin{equation}
(U^{CB} \mathcal{M}  )_{\hat{a} \hat{b} }^{ab} \equiv (\bar{\Psi}^{B}X^{C})_{c}^{a}   \mathcal{M}_{\hat{a} \hat{b} }^{cb}+(\bar{\Psi}^{B}X^{C})_{c}^{b}   \mathcal{M}_{\hat{a} \hat{b} }^{ac}- \mathcal{M}_{\hat{c} \hat{b} }^{ab} (X^{C} \bar{\Psi}^{B})^{\hat{c}}_{\hat{a}}- \mathcal{M}_{\hat{a} \hat{c} }^{ab} (X^{C} \bar{\Psi}^{B})^{\hat{c}}_{\hat{b}}\;,
\end{equation}
respectively. The current conservation equation $ \partial^{\mu}j^{AB}_{\mu }=0 $ requires 
\begin{equation}\label{89i}
D^{\mu}\mathcal{M}=0\;,\;\;\;\;\;  W^{C}_{D} \mathcal{M}  =0   \;,\;\;\;\;\; V_{CD} \mathcal{M}=0  \;,\;\;\;\;\;U^{CD} \mathcal{M} =0\;,
\end{equation}
$ \forall \; C,D=1,\cdots,4 $. (\ref{89i}) is obtained from the equations of motion and the algebraic calculations with no explicit form for monopole operators assumed. In the following, we will show that the constructed monopole operators satisfy (\ref{89i}).

In section \ref{der}, $ D^{\mu}\mathcal{M}=0 $ has been proven. $ V_{CD} \mathcal{M}=0 $ and $U^{CD} \mathcal{M} =0  $ are related to $ W^{C}_{D} \mathcal{M}  =0 $ by a $\mathcal{N}=6  $ supersymmetry transformation. Suppose $ \Gamma^{I}_{AB} \varepsilon^{I}=\eta_{AB}  $, (\ref{817a})-(\ref{817b}) could be written as 
\begin{eqnarray}
  \label{911aa}&&  \delta X_{A}=i  \bar{\Psi}^{B}  \eta_{AB}   \;,  \\ \label{911aaa} &&   \delta \Psi_{A}=\gamma^{\mu} \eta_{AB}  D_{\mu}X^{B}+\eta_{AB}(X^{C}X_{C}X^{B}-X^{B}X_{C}X^{C})-2\eta_{BC}X^{B}X_{A}X^{C}\;,\\ &&  \delta A_{\mu }=\frac{1}{2}\epsilon^{ABCD}
 X_{B} \bar{\Psi}_{A} \gamma_{\mu}\eta_{CD}+\bar{\Psi}^{A}  \gamma_{\mu} \eta_{AB} X^{B}\;,\\ &&  \delta \hat{A}_{\mu}=\frac{1}{2}\epsilon^{ABCD}\bar{\Psi}_{A} \gamma_{\mu}\eta_{CD}X_{B}+X^{B} \bar{\Psi}_{A}  \gamma_{\mu}\eta_{AB}\;.
\end{eqnarray}
The supersymmetry transformations of $ X^{C}X_{D} $, $ X_{D}X^{C} $ and $ \mathcal{M} $ are
\begin{eqnarray}
&&\delta (X^{C}X_{D})=-\frac{i}{2}\epsilon^{C ABE} \bar{\eta}_{BE}  \Psi_{A}X_{D}+iX^{C} \bar{\Psi}^{B}  \eta_{DB}\;, \\ && \delta (X_{D}X^{C})=-\frac{i}{2}\epsilon^{C ABE}  \bar{\eta}_{BE} X_{D} \Psi_{A}+i \bar{\Psi}^{B} X^{C} \eta_{DB}\;, \\ &&  \delta \mathcal{M}=0   \;. 
\end{eqnarray}
For the given $V_{CD}  $ or $U^{CD}   $, one can always find the suitable $W^{A}_{B}  $ and $ \eta $ with 
\begin{equation}
\delta (W^{A}_{B}  \mathcal{M})=i  (V_{CD} \mathcal{M})\;,
\end{equation}
or 
\begin{equation}
\delta (W^{A}_{B}  \mathcal{M})=i  (U^{CD}  \mathcal{M})\;.
\end{equation}
So
\begin{equation}
W^{A}_{B}  \mathcal{M}  =0 \Rightarrow V_{CD} \mathcal{M}=0\;, \;\;\;\;\;W^{A}_{B} \mathcal{M}  =0 \Rightarrow U^{CD} \mathcal{M} =0\;.
\end{equation}

It remains to show $ W^{C}_{D} \mathcal{M}  =0 $. The operator realization of $  W^{C}_{D} $ is $ -i \mathcal{K}^{C}_{D} $ with  
\begin{eqnarray}\label{kcd1}
\mathcal{K}^{C}_{D} \nonumber    &=&\int d^{2} x \;  tr [(\Pi^{B}X_{B}-X^{B}\Pi_{B} )X^{C}X_{D}+(\Pi_{B}X^{B} -X_{B}\Pi^{B})X_{D}X^{C} \\ &-& \frac{  k}{2\pi}\epsilon^{ij}D_{i}X^{C}D_{j}X_{D}-\frac{ i k}{2\pi}(
\bar{\Psi}^{B}\gamma_{0}\Psi_{B} 
X_{D}X^{C}   +\bar{\Psi}_{B}\gamma_{0}\Psi^{B}  X^{C}X_{D}  
 ) ]  \;.
\end{eqnarray}
For example, 
\begin{equation}
-i[ \mathcal{K}^{C}_{D},X^{B}]=  X^{C}X_{D}X^{B}-X^{B}X_{D}X^{C}\;,\;\;\;\;\;\;\;\;
-i[ \mathcal{K}^{C}_{D},A_{i}]=-iD_{i}(X_{D}X^{C})\;.
\end{equation}
$ \mathcal{K}^{C}_{D} $ is gauge invariant and does not contain $ F $, so $ M_{R}(x)  \mathcal{K}^{C}_{D}M^{-1}_{R}(x) =0$, $ [M_{R}(x) , \mathcal{K}^{C}_{D}]=0 $, 
\begin{equation}
W^{C}_{D} \mathcal{M}= -i[ \mathcal{K}^{C}_{D},\mathcal{M}]=0 \;.
\end{equation}
Up to now, (\ref{89i}) is verified.

In ABJM theory, from the SUSY invariance of the monopole operator, we can arrive at a conclusion similar to (\ref{89i}). In superspace formulation of the gauge theory, commutators of the covariant derivatives in superspace yield the field strengths, on which the constraints can be imposed. Supersymmetric gauge theories are entirely characterized by these constraints \cite{22b}. For ABJM theory in $ \mathcal{N} =6$ superspace parametrized by coordinates $(x^{\mu},\theta^{I})  $, covariant derivatives should satisfy the constraints \cite{tw13}
\begin{equation}\label{820a}
\{D^{I},\bar{D}^{J}\}=2i\delta^{IJ}\gamma^{\mu}D_{\mu}+i \mathcal{F}^{IJ}\;,
\end{equation}
where $ D^{I} $ is the covariant derivative along $\theta^{ I}  $, $ D_{\mu} $ is the covariant derivative in 3d spacetime, and $ \mathcal{F}^{IJ}=-\mathcal{F}^{JI} $ is the $ IJ $ component of the field strength in superspace whose lowest $ \theta $-expansion gives $ \Sigma^{IJ \;D}_{C}(X^{C}X_{D})^{\hat{a}}_{\hat{b}}$ and $ \Sigma^{IJ \;D}_{C}(X_{D}X^{C})^{a}_{b} $.
\begin{equation}
 \Sigma^{IJ \;D}_{C}=\Gamma^{[I}_{CA}\tilde{\Gamma}^{J]AD} \;,\;\;\;\;\;\;\; (\Sigma^{IJ \;D}_{C})^{*}=-\Sigma^{IJ \;C}_{D} \;.
\end{equation}
$ \Sigma^{IJ \;C}_{C}=  0$, so effectively, 
\begin{equation}
 \Sigma^{IJ \;D}_{C}  X^{C}X_{D} \sim X^{C}X_{D} -\frac{1}{4} \delta^{C}_{D}X^{B}X_{B}\;,\;\;\;\;\;\;\;
 \Sigma^{IJ \;D}_{C}X_{D}X^{C} \sim X_{D}X^{C}-\frac{1}{4} \delta^{C}_{D}X_{B}X^{B}\;.
\end{equation}

The operator realization of $D^{I}$ and $\bar{D}^{J}  $ is $ \mathcal{Q}^{I}$ and $ \bar{\mathcal{Q}}^{J} $. One may compute the anticommutator of supercharges directly, 
\begin{equation}\label{824e}
\{\mathcal{Q}^{I}, \bar{\mathcal{Q}}^{J}\}  =-2\delta^{IJ}\gamma^{\mu}\mathcal{P}_{\mu}+
 \Sigma^{IJ \;D}_{C}\mathcal{K}^{C}_{D} \;.
\end{equation}
$ \mathcal{P}_{\mu} $ is the gauge invariant energy momentum operator of ABJM theory which acts on fields as $ -iD_{\mu} $. With $\mathcal{K}^{C}_{D}  $ given by (\ref{kcd1}), the action of $\Sigma^{IJ \;D}_{C}\mathcal{K}^{C}_{D}  $ is the gauge variation $i \Sigma^{IJ \;D}_{C}W^{C}_{D}  $. When acting on gauge invariant operators, (\ref{824e}) reduces to the standard supersymmetry algebra 
\begin{equation}
\{\mathcal{Q}^{I}, \bar{\mathcal{Q}}^{J}\} = -2\delta^{IJ}\gamma^{\mu}P_{\mu}
\end{equation}
with $ P_{\mu} $ the energy momentum operator. $[\mathcal{Q}^{I} ,\mathcal{M}  ]=0$, so $[\{\mathcal{Q}^{I} ,\bar{\mathcal{Q}}^{J}\},\mathcal{M}  ]=0  $. From (\ref{824e}), when $ I=J $, we get $ D_{\mu} \mathcal{M}=0$; when $ I \neq J $, we have $\Sigma^{IJ \;D}_{C}W^{C}_{D}   \mathcal{M}=0$, which is actually a weaker statement compared with $ W^{C}_{D}   \mathcal{M}=0 $.

The anticommutator of $ \mathcal{Q}^{I}$ and $\bar{\mathcal{S}}^{J}$ gives more superconformal charges. For example, from $ \{ \mathcal{Q}^{I}, \bar{\mathcal{S}}^{I}\} $, we may get $ \mathcal{D} $, which is the gauge invariant completion of the dilation $ D $ and would reduce to $ D $ when acting on gauge invariants. Since $ [ \mathcal{Q}^{I}, \mathcal{M}_{R} ]=[\mathcal{S}^{J} , \mathcal{M}_{R} ]=0 $, $ [\mathcal{D} ,  \mathcal{M}_{R}]=0 $. $  \mathcal{M}_{R} $ have the conformal dimension $ 0 $. In section \ref{cla}, for $\mathcal{M}_{R}  $ to have the conformal dimension $ 0 $ under the action of $ D $, $ a_{i} $ must take the special form (\ref{55f}) which is not necessary when considering the gauge invariants.

Usually, the current conservation equation may impose some constraints on the classical Lagrangian. For example, $\partial^{\mu}  j^{A}_{\mu B} =0 $ requires that the Lagrangian must be $ SU(4) $ invariant. To study the constraints imposed by $\partial^{\mu} j^{AB}_{\mu } =0 $, consider a truncated ABJM model with no spinor fields for simplicity, where the monopole operators satisfy $D_{\mu} \mathcal{M}=0  $ and $ W^{C}_{D}   \mathcal{M}=0 $. Among all of the $ SU(4) $ invariant scalar potentials, only the mass term $ m^{2} tr(X^{A}X_{A})$ and the sextic potential in (\ref{613i}) could make $ j^{AB}_{\mu }  $ conserved. The requirement of symmetry enhancement almost fixes the interaction potential.

From the conserved currents (\ref{jab1}) and (\ref{jab}), $15+6+6= 27 $ conserved R-symmetry charges 
\begin{eqnarray}
\nonumber R^{A}_{ B} &=&-\frac{ik}{2\pi}\int d^{2}y \;\{ [X_{a}^{A \hat{a}}(D_{0}X_{B})^{a}_{\hat{a}}-(D_{0}X^{A})_{a}^{\hat{a}}X_{B\hat{a}}^{a}+i \bar{\Psi}^{Aa}_{\hat{a}}\gamma_{0} \Psi_{Ba}^{ \hat{a}} ]\\ &-&\frac{1}{4}\delta_{B}^{A} [X_{a}^{C \hat{a}}(D_{0}X_{C})^{a}_{\hat{a}}-(D_{0}X^{C})_{a}^{\hat{a}}X_{C\hat{a}}^{a}+i \bar{\Psi}^{Ca}_{\hat{a}}\gamma_{0} \Psi_{Ca}^{ \hat{a}} ]\}   \;,
\end{eqnarray}
\begin{equation}
R^{AB}= -\frac{ik}{4\pi}\int d^{2}y \;\mathcal{M}_{\hat{a} \hat{b} }^{ab}[X_{a}^{A \hat{a}}(D_{0}X^{B})_{b}^{\hat{b}}-(D_{0}X^{A})_{a}^{\hat{a}}X_{b}^{B \hat{b}}+\frac{i}{2}\epsilon^{ABCD} \bar{\Psi}_{Ca}^{\hat{a}}\gamma_{0} \Psi_{Db}^{ \hat{b}}]
\end{equation}
and
\begin{equation}
R_{AB}=R^{AB+}= \frac{ik}{4\pi}\int d^{2}y \;\mathcal{M}^{\hat{a} \hat{b} }_{ab}[X^{a}_{A \hat{a}}(D_{0}X_{B})^{b}_{\hat{b}}-(D_{0}X_{A})^{a}_{\hat{a}}X^{b}_{B \hat{b}}-\frac{i}{2}\epsilon_{ABCD} \bar{\Psi}^{Ca}_{\hat{a}}\gamma_{0} \Psi^{Db}_{ \hat{b}}]
\end{equation}
are obtained, which, together with (\ref{12qa11}), compose the generators of the $ SO(8) $ group. The commutation relations of the R-symmetry charges obey the $ so(8) $ algebra. In particular, 
\begin{equation}
[R^{AB}, R_{CD}]=  \frac{1}{4}[\delta_{D}^{A}R^{B}_{ C}  -\delta_{C}^{A}  R^{B}_{ D}
+   \delta_{C}^{B}R^{A}_{ D}  - \delta_{D}^{B} R^{A}_{ C} ]
\end{equation}
requires $ \mathcal{M}_{\hat{a} \hat{b} }^{ab}\mathcal{M}^{\hat{c} \hat{b} }_{cb}= \delta^{a}_{c }\delta^{\hat{c} }_{\hat{a}}$, which is proved in (\ref{427f}).

\section{$ \mathcal{N}=8 $ supersymmetry in ABJM theory}\label{SO8}

The dynamical fields of the ABJM theory consist of $X^{a}_{A \hat{a}}  $, $\Psi_{Aa}^{ \hat{a}}    $ in $ \bar{4} $ representation and the adjoints $ X_{a}^{A \hat{a}} $, $  \Psi^{Aa}_{ \hat{a}} $ in $ 4 $ representation of the $ SU(4)$ R-symmetry. Although there are $ 8 $ scalars $(X_{A},X^{A})  $ and $ 8 $ spinors $ (\Psi_{A},\Psi^{A}) $, they cannot be transformed into each other by the $ SO(8) $ rotations due to the distinct gauge group representations. With the nondynamical Chern-Simons gauge fields $ A^{a}_{\mu b} $ and $\hat{A}^{\hat{a}}_{\mu \hat{b}}  $ added, the monopole operator $ \mathcal{M}_{R }(x) $ can be obtained. In particular, when $ k=1,2 $, using $ \mathcal{M}_{\hat{a} \hat{b} }^{ab}(x)   $ and $  \mathcal{M}^{\hat{a} \hat{b} }_{ab}(x)  $, one may construct the dressed fields \cite{tw10,tw11}
\begin{equation}\label{91q}
\tilde{X}^{A a}_{ \hat{a}} (x)=  \mathcal{M}_{\hat{a} \hat{b} }^{ab}(x) X_{b}^{A \hat{b}}(x)\;,\;\;\;\;\;\;\;\tilde{\Psi}^{ a}_{A\hat{a}} (x)=  \mathcal{M}_{\hat{a} \hat{b} }^{ab}(x) \Psi_{Ab}^{ \hat{b}}(x)
\end{equation}
with the adjoints
\begin{equation}\label{92q}
\tilde{X}_{A a}^{ \hat{a}} (x)=  \mathcal{M}^{\hat{a} \hat{b} }_{ab}(x) X^{b}_{A \hat{b}}(x)\;,\;\;\;\;\;\;\;\tilde{\Psi}_{ a}^{A\hat{a}} (x)=  \mathcal{M}^{\hat{a} \hat{b} }_{ab}(x) \Psi^{Ab}_{ \hat{b}}(x)\;.
\end{equation}
The $ 8 $ scalars $(X_{A},\tilde{X}^{A})  $ and the $ 8 $ spinors $ (\tilde{\Psi}_{A},\Psi^{A}) $ are now in the same representation of the gauge group, and thus could be transformed into each other by the $ SO(8) $ rotations.

Except for modifying the gauge group representation, $\mathcal{M}_{R} (x) $ are ``silent" in several respects. They have the weight $ 0 $ and thus would not change the dressed operator's conformal dimension; they commute with the supercharges and thus would not influence the dressed operator's supersymmetry; they are covariantly constant and thus would not affect the dressed operator's covariant derivative. This is different from the monopole operator in 3d Yang-Mills-matter theories where the gauge fields are dynamical.

The $ SO(8) $ symmetry group has $ 28=15+6\cdot 2 +1$ real parameters. The first $ 15 $ are $ \omega^{B}_{A} $'s satisfying $ (\omega^{A}_{B} )^{*}+\omega^{B}_{A}=0  $, $ \omega^{A}_{A}=0 $. $ \omega^{B}_{A}R^{A}_{B} $ generates the $ SU(4) $ transformation, under which 
\begin{equation}
\delta X_{A }=\omega^{B}_{A}X_{B }\;,\;\;\;\;\;\; \delta \tilde{X}^{A}=-\omega^{A}_{B}\tilde{X}^{B}\;,\;\;\;\;\;\;\delta  \tilde{\Psi}_{A} = \omega^{B}_{A}\tilde{\Psi}_{B}  \;,\;\;\;\;\;\;\delta \Psi^{A}  =-\omega_{B}^{A}\Psi^{B}\;,
\end{equation}
and 
\begin{equation}
\delta X^{A }=-\omega_{B}^{A}X^{B }\;,\;\;\;\;\;\; \delta \tilde{X}_{A}=\omega_{A}^{B}\tilde{X}_{B}\;,\;\;\;\;\;\;\delta  \tilde{\Psi}^{A }= -\omega^{A}_{B}\tilde{\Psi}^{B }  \;,\;\;\;\;\;\;\delta \Psi_{A}= \omega^{B}_{A}\Psi_{B}
\end{equation}
for adjoints. The $ 6\cdot 2 $ real parameters are $ \omega^{AB} $'s satisfying 
\begin{equation}
 \omega^{AB}+ \omega^{BA}=0\;,\;\;\;\;\;\; (\omega^{AB})^{*}+ \omega_{BA}=0\;.
\end{equation}
$ R_{AB} \omega^{AB}$ and $ R^{AB} \omega_{AB} $ generate the transformations 
\begin{equation}\label{96d}
\delta X_{A }=\omega_{AB}\tilde{X}^{B}\;,\;\;\;\;\;\; \delta \tilde{X}^{A } =\omega^{AB}X_{ B}\;,\;\;\;\;\;\;\delta  \tilde{\Psi}_{A} = \frac{1}{2}\epsilon_{ABCD}\omega^{BC}\Psi^{ D}  \;,\;\;\;\;\;\;\delta \Psi^{A} =\frac{1}{2}\epsilon^{ABCD}\omega_{BC}\tilde{\Psi}_{ D}
\;,
\end{equation}
and 
\begin{equation}\label{97d}
\delta X^{A }=\omega^{AB}\tilde{X}_{B}\;,\;\;\;\;\;\; \delta \tilde{X}_{A } =\omega_{AB}X^{B } \;,\;\;\;\;\;\;\delta  \tilde{\Psi}^{A }=\frac{1}{2}\epsilon^{ABCD}\omega_{BC}\Psi_{ D}  \;,\;\;\;\;\;\;\delta \Psi_{A} = \frac{1}{2}\epsilon_{ABCD}\omega^{BC}\tilde{\Psi}^{ D}
\end{equation}
for adjoints. For (\ref{96d}) and (\ref{97d}) to be consistent with (\ref{91q}) and (\ref{92q}), 
\begin{equation}
\mathcal{M}_{\hat{a} \hat{b} }^{ab}(x)\mathcal{M}^{\hat{a} \hat{c} }_{ac}(x) = \delta^{b}_{c}  \delta^{\hat{c} }_{\hat{b} }
\end{equation}
is again needed.

Aside from the monopole dressed matter fields, one may also introduce the monopole dressed gauge fields $  \tilde{A}^{c}_{\mu a} $ and $  \tilde{\hat{A}}^{\hat{c}}_{\mu \hat{a}}$ with  
\begin{equation}\label{99i}
 \tilde{A}^{a}_{\mu c}=\mathcal{M}_{cd} A^{d}_{\mu b}\mathcal{M}^{ba}+i \mathcal{M}_{cb}   \partial_{\mu}\mathcal{M}^{ba}\;,\;\;\;\;\;\;\;\;
\tilde{\hat{A}}^{\hat{c}}_{\mu \hat{a}}=\mathcal{M}_{\hat{a} \hat{b} }\hat{A}^{\hat{b}}_{\mu \hat{d}} \mathcal{M}^{\hat{d} \hat{c} }+i \mathcal{M}_{\hat{a} \hat{b} }\partial_{\mu} \mathcal{M}^{\hat{b} \hat{c} } 
\end{equation}
so that 
\begin{equation}\label{913}
(\tilde{D}_{\mu} \tilde{X}^{B})^{a}_{ \hat{a}} = \partial_{\mu}  \tilde{X}^{Ba}_{ \hat{a}} 
+i(\tilde{A}^{a}_{\mu c} \tilde{X}^{Bc}_{\hat{a}}-\tilde{X}^{Ba}_{\hat{c}}\tilde{\hat{A}}^{\hat{c}}_{\mu \hat{a}})=\mathcal{M}^{ab}_{\hat{a} \hat{b}}(D_{\mu} X^{B})_{b}^{ \hat{b}}     \;,
\end{equation}
\begin{equation}\label{914}
(\tilde{D}_{\mu} \tilde{X}_{B})_{a}^{ \hat{a}} = \partial_{\mu}  \tilde{X}_{Ba}^{ \hat{a}} 
+i(\tilde{\hat{A}}_{\mu \hat{c}}^{\hat{a}}\tilde{X}_{Ba}^{\hat{c}}- \tilde{X}_{Bc}^{\hat{a}}\tilde{A}_{\mu a}^{ c}) =\mathcal{M}_{ab}^{\hat{a} \hat{b}}(D_{\mu} X_{B})^{b}_{ \hat{b}}      \;,
\end{equation}
where we have used 
\begin{equation}
\mathcal{M}^{ab}_{\hat{a} \hat{b}}=\mathcal{M}^{ab}\mathcal{M}_{\hat{a} \hat{b}}\;,\;\;\;\;\;\;\;\;\mathcal{M}_{ab}^{\hat{a} \hat{b}}=\mathcal{M}_{ab}\mathcal{M}^{\hat{a} \hat{b}}\;,\;\;\;\;\;\;\;\;
\mathcal{M}_{ac}\mathcal{M}^{cb}=\delta^{b}_{a}\;,\;\;\;\;\;\;\;\;\mathcal{M}_{\hat{a} \hat{c}}\mathcal{M}^{\hat{c} \hat{b}}=\delta^{ \hat{b}}_{ \hat{a}}\;.
\end{equation}
With the fields replaced by the monopole dressed fields, the ABJM action (\ref{69i}) is invariant: 
\begin{equation}\label{99s}
S(X^{a}_{A\hat{a}},X_{a}^{A\hat{a}},\Psi^{\hat{a}}_{Aa},\Psi_{\hat{a}}^{Aa}, A_{b}^{a}, \hat{A}^{\hat{a}}_{\hat{b}})=S(\tilde{X}^{\hat{a}}_{Aa},\tilde{X}_{\hat{a}}^{Aa},\tilde{\Psi}^{a}_{A\hat{a}},\tilde{\Psi}_{a}^{A\hat{a}}, \tilde{A}^{b}_{a}, \tilde{\hat{A}}_{\hat{a}}^{\hat{b}})\;.
\end{equation}
Moreover, $ D_{\mu}\mathcal{M}^{ba}= D_{\mu} \mathcal{M}^{\hat{b} \hat{c} }=0$ is equivalent to the condition 
\begin{equation}
\tilde{A}^{a}_{\mu c}= -A^{a}_{\mu c}\;,\;\;\;\;\;\;\;\tilde{\hat{A}}^{\hat{c}}_{\mu \hat{a}}=- \hat{A}^{\hat{c}}_{\mu \hat{a}}\;,
\end{equation}
so the gauge fields are invariant under the monopole dressing if and only if the monopole operators are covariantly constant.

With the R-charges $ R^{AB} $ and $ R_{AB} $ added into the $ \mathcal{N} =6$ superconformal algebra, two additional supercharges will be generated. For convenience, we will use 
\begin{equation}
\mathcal{R}^{I} =\Gamma_{AB}^{I}R^{AB} \;,\;\;\;\;\;\;\;\;\;  \mathcal{R}^{I+} =-\tilde{\Gamma}^{IAB}R_{AB}  
\end{equation}
instead of $ R^{AB} $ and $ R_{AB} $. The commutator of $ \mathcal{Q}^{I}$ and $\mathcal{R}^{J} $ gives two extra supercharges $ \mathcal{Q} $ and the adjoint $ \mathcal{Q}^{+} $:
\begin{equation}\label{916g}
[\mathcal{R}^{J},\mathcal{Q}^{I}]=\delta^{IJ} \mathcal{Q} \;,\;\;\;\;\;\;\;\;\;[\mathcal{R}^{J+},\mathcal{Q}^{I}]=-\delta^{IJ}\mathcal{Q}^{+} \;,
\end{equation}
where 
\begin{equation}\label{917g}
 \mathcal{Q}=-\frac{k}{2\pi} \int d^{2}x \;tr[(2 \gamma^{\mu}D_{\mu}\tilde{X}^{D}  +X_{A}X^{A }\tilde{X}^{D}- \tilde{X}^{D}X^{A }X_{A}     )\gamma_{0}\Psi_{D}+\frac{4}{3}\epsilon_{ABCD}X^{A}\tilde{X}^{B }X^{C}
\gamma_{0}\Psi^{D}]\;.
\end{equation}
To get (\ref{916g}) and (\ref{917g}), we have used the properties (\ref{89i}) and $ [\mathcal{Q}^{I}, \mathcal{M}]=0 $. Also, because of (\ref{89i}), the right-hand side of (\ref{917g}) can take several different but equivalent forms.

For example, using $ U^{CD} \mathcal{M}=0$, we have
\begin{eqnarray}
   &&  tr[(X^{B}\tilde{X}^{A }X^{C}-X^{C}\tilde{X}^{A }X^{B})\gamma_{0}\Psi^{M}] =   tr[(X^{C}\tilde{X}^{B }X^{A }-X^{A }\tilde{X}^{B }X^{C})\gamma_{0}\Psi^{M}]\\ &=& tr[(X^{A }\tilde{X}^{C }X^{B}-X^{B}\tilde{X}^{C }X^{A })\gamma_{0}\Psi^{M}]= \frac{1}{3}\epsilon^{BACM}\epsilon_{PQRM}tr[(X^{P }\tilde{X}^{Q }X^{R})\gamma_{0}\Psi^{M}]   \;,
\end{eqnarray}
so effectively, 
\begin{equation}
X^{B}\tilde{X}^{A }X^{C}-X^{C}\tilde{X}^{A }X^{B}\sim \frac{1}{3}\epsilon^{BACM}\epsilon_{PQRM}(X^{P }\tilde{X}^{Q }X^{R}) \sim \bar{4}
\end{equation}
is in $ \bar{4} $ representation of $ SU(4) $. In \cite{tw13}, the constraint $ X^{B}\tilde{X}^{A }X^{C}-X^{C}\tilde{X}^{A }X^{B}\sim \bar{4} $ was derived in Eq (6.11) as a consequence of $ D_{\mu} \mathcal{M}=0$. We have seen that such a condition can indeed be satisfied.

$  \mathcal{Q} $, $\mathcal{Q}^{+}  $, and $  \mathcal{Q}^{I}  $ altogether comprise $ 8 $ supercharges for the $ \mathcal{N} =8$ supersymmetry. Under the action of $\bar{\mathcal{Q} }  \varepsilon$, 
\begin{eqnarray}
   && \label{1021} \delta X_{D}=2i \bar{\tilde{\Psi}}_{D}\varepsilon  \;,\;\;\;\;\;\;\;\;\;  \delta X^{D}=0 \;, \\  &&   \delta \Psi^{D}=-2\gamma^{\mu}
  D_{\mu}\tilde{X}^{D} \varepsilon +(X_{A}X^{A }\tilde{X}^{D}- \tilde{X}^{D}X^{A }X_{A}     )\varepsilon\;,\\ && \delta \Psi_{D}=-\frac{4}{3}\epsilon_{ABCD}X^{A}\tilde{X}^{B }X^{C}\varepsilon\;,\\  && \delta A_{i}= 2 \tilde{X}^{D}\bar{\Psi}_{D}\gamma_{i} \varepsilon \;,\;\;\;\;\;\;\;\;\;\delta \hat{A}_{i}=- 2\bar{\Psi}_{D} \gamma_{i} \varepsilon \tilde{X}^{D}   \;,
\end{eqnarray}
and similarly for $ \bar{\mathcal{Q}}^{+}  \varepsilon^{+}  $. In (\ref{917g}), $  \mathcal{Q} $ and $\mathcal{Q}^{+}  $ do not contain $ F $, and thus should commute with $ M_{R} (x)$. $ [ \mathcal{Q}, M_{R} (x)]= [ \mathcal{Q}^{+}, M_{R} (x)] =0$, $ [ \mathcal{Q}, \mathcal{M}_{R} (x)]= [ \mathcal{Q}^{+}, \mathcal{M}_{R} (x)] =0$, $  \mathcal{M}_{R} (x) $ is invariant under the $ \mathcal{N} =8$ supersymmetry.

From (\ref{824e}), (\ref{916g}), and the $ so(8) $ algebra, we also have 
\begin{equation}
[\mathcal{R}^{I}, \mathcal{Q} ]=0\;,\;\;\;\;\;\;\;[\mathcal{R}^{I+},\mathcal{Q} ]=-2\mathcal{Q}^{I}\;,\;\;\;\;\;\;\;[\mathcal{R}^{I+}, \mathcal{Q}^{+} ]=0\;,\;\;\;\;\;\;\;[\mathcal{R}^{I},\mathcal{Q}^{+} ]=2\mathcal{Q}^{I}\;, 
\end{equation}
\begin{equation}\label{919a}
\{\mathcal{Q}, \bar{\mathcal{Q}}^{I}\} =\mathcal{K}^{I }\;,\;\;\;\;\;\;\;\{\mathcal{Q}^{+}, \bar{\mathcal{Q}}^{I}\} =\mathcal{K}^{I +}\;,\;\;\;\;\;\;\; \{\mathcal{Q}, \bar{\mathcal{Q}}^{+}\} =4\gamma^{\mu}\mathcal{P}_{\mu}+\mathcal{K}\;,
\end{equation}
where $ \mathcal{K}^{I } $ is a gauge variation with $ 6+6 $ parameters $ 2\Gamma^{I}_{MC} \tilde{X}^{M }X^{C} $ and $ 2\Gamma^{I}_{MC} X^{C}\tilde{X}^{M }$,
\begin{equation}
[\mathcal{K}^{I },X_{A}]=2i \Gamma^{I}_{MC}( \tilde{X}^{M }X^{C}X_{A}-X_{A}X^{C}\tilde{X}^{M })\;;
\end{equation}
$ \mathcal{K}^{I +} $ is a gauge variation with $ 6+6 $ parameters $ -2\tilde{\Gamma}^{IMC}X_{M}\tilde{X}_{C }$ and $ -2\tilde{\Gamma}^{IMC} \tilde{X}_{C }X_{M}$,
\begin{equation}
[\mathcal{K}^{I +},X_{A}]=-2i \tilde{\Gamma}^{IMC}( X_{M}\tilde{X}_{C }X_{A}-X_{A}\tilde{X}_{C }X_{M})\;;
\end{equation}
$ \mathcal{K} $ is a gauge variation with $ 1+1 $ parameters $ 2(X_{C }X^{C}-\tilde{X}^{C }\tilde{X}_{C } )$ and $2(X^{C}X_{C }-\tilde{X}_{C }\tilde{X}^{C }  )$, 
\begin{equation}
[\mathcal{K},X_{A}]
=2i  [(X_{C }X^{C}-\tilde{X}^{C }\tilde{X}_{C })X_{A}-X_{A}(X^{C}X_{C }-\tilde{X}_{C }\tilde{X}^{C })]\;.
\end{equation}
When acting on the gauge invariant operators, (\ref{919a}) becomes 
\begin{equation}
\{\mathcal{Q}, \bar{\mathcal{Q}}^{I}\} =0\;,\;\;\;\;\;\;\;\{\mathcal{Q}^{+}, \bar{\mathcal{Q}}^{I}\} =0\;,\;\;\;\;\;\;\;\{\mathcal{Q}, \bar{\mathcal{Q}}^{+}\} =4\gamma^{\mu}P_{\mu}\;,
\end{equation}
completing the $ \mathcal{N} =8$ superalgebra.

In \cite{tw15}, a systematic classification of the unitary superconformal multiplets in $ d \geq 3 $ spacetime dimensions is given. In the following, specified to ABJM model, we will give the explicit operator content for the $ 1/3$ BPS stress tensor multiplet, $1/2$ BPS extra SUSY-current multiplet, $1/6$ BPS higher-spin current multiplet, free hypermultiplet of the 3d $  \mathcal{N} =6 $ SCFT and the $1/2 $ BPS stress tensor multiplet of the 3d $  \mathcal{N} =8 $ SCFT. The operators are characterized by $ [2j]_{\Delta}^{(R)}   $, where $ j $ is the half-integer $ su(2) $ spin, $ \Delta $ is the conformal dimension, and $ R $ is the Dynkin labels of the R-symmetry representation \cite{tw15}.

We will take
\begin{equation}
\bar{\mathcal{Q}}^{AB}=-\frac{1}{4}\bar{Q}^{I}  \tilde{\Gamma}^{IAB} 
\end{equation}
instead of $\bar{\mathcal{Q}}^{I}  $ for $  \mathcal{N} =6 $ supercharges. Then 
\begin{eqnarray}
   && \label{1033} [\bar{\mathcal{Q}}^{AB} \eta_{AB} ,X_{C}]=i  \bar{\Psi}^{B}  \eta_{CB}  \;,\;\;\;\;\;
[\bar{\mathcal{Q}}^{AB} \eta_{AB} ,X^{C}]=-\frac{i}{2}\epsilon^{CDMN}\bar{\Psi}_{D}\eta_{MN}\;,  \\ \label{1033d}   &&   [\bar{\mathcal{Q}}^{AB} \eta_{AB} ,\tilde{X}_{C}]=i  \bar{\tilde{\Psi}}^{B}  \eta_{CB}  \;,\;\;\;\;\;
[\bar{\mathcal{Q}}^{AB} \eta_{AB} ,\tilde{X}^{C}]=-\frac{i}{2}\epsilon^{CDMN}\bar{\tilde{\Psi}}_{D}\eta_{MN}\;,\\ \label{1033de} && [ \bar{\mathcal{Q}} ,X^{C}]=0\;,\;\;\;\;\;[ \bar{\mathcal{Q}} ,\tilde{X}^{C}]=0\;,\;\;\;\;\;[ \bar{\mathcal{Q}}^{+} ,X_{C}]=0\;,\;\;\;\;\;[ \bar{\mathcal{Q}}^{+} ,\tilde{X}_{C}]=0\;,
\end{eqnarray}
where $ [\bar{\mathcal{Q}}^{AB}  , \mathcal{M}]= [ \bar{\mathcal{Q}} ,\mathcal{M}]=[ \bar{\mathcal{Q}}^{+} ,\mathcal{M}]= 0$ is used.

According to \cite{tw15}, in 3d $  \mathcal{N} =6 $ SCFT, there is a protected $ 1/3 $ BPS stress tensor multiplet with chiral primary operators in $ [0]^{(0,1,1)}_{1} $ representation. Successive action of the supercharges gives the complete $ 64+64 $ multiplet:
\begin{equation}
[0]^{(0,1,1)}_{1}\stackrel{Q}{\longrightarrow}[1]^{(0,0,2)\oplus (0,2,0)   \oplus  (1,0,0)     }_{\frac{3}{2}}\stackrel{Q}{\longrightarrow} [0]^{(0,1,1)}_{2}\oplus [2]^{(0,0,0)\oplus (0,1,1)}_{2}\stackrel{Q}{\longrightarrow}[3]^{(1,0,0)}_{\frac{5}{2}}\stackrel{Q}{\longrightarrow}[4]^{(0,0,0)}_{3}\;.
\end{equation}
In ABJM theory, $ [0]^{(0,1,1)}_{1} $ are $ 15 $ scalars $ tr(X^{A}X_{B}-\frac{1}{4}\delta_{B}^{A}X^{C}X_{C}) $ annihilated by $ 2 $ supercharges. The action of the rest $ 4 $ supercharges gives the whole multiplet. The $ 1+15 $ dimension-$ 2 $ vectors $  [2]^{(0,0,0)\oplus (0,1,1)}_{2} $ are $ U(1)_{J} $ current and the $ SU(4) $ R-symmetry current $j^{A}_{\mu B}  $ in (\ref{jab1}). $ [3]^{(1,0,0)}_{\frac{5}{2}} $ is the SUSY current whose integration gives the supercharges $ \mathcal{Q}^{I} $. $ [4]^{(0,0,0)}_{3} $ is the stress tensor.

Sometimes, the theory also contains $ 32+32 $ $ 1/2 $ BPS extra SUSY-current multiplet with chiral primary operators in $[0]^{(0,0,2)}_{1}$ representation. The whole multiplet is given by 
\begin{equation}\label{1026}
[0]^{(0,0,2)}_{1}\stackrel{Q}{\longrightarrow}[1]^{(0,1,1)}_{\frac{3}{2}}\stackrel{Q}{\longrightarrow} [0]^{(0,2,0)}_{2}\oplus [2]^{(1,0,0)}_{2}\stackrel{Q}{\longrightarrow}[3]^{(0,0,0)}_{\frac{5}{2}}
\end{equation}
with the complex conjugation
\begin{equation}
[0]^{(0,2,0)}_{1}\stackrel{Q}{\longrightarrow}[1]^{(0,1,1)}_{\frac{3}{2}}\stackrel{Q}{\longrightarrow} [0]^{(0,0,2)}_{2}\oplus [2]^{(1,0,0)}_{2}\stackrel{Q}{\longrightarrow}[3]^{(0,0,0)}_{\frac{5}{2}}\;.
\end{equation}
In ABJM theory with $ k=1,2 $, $[0]^{(0,0,2)}_{1}$ and $ [0]^{(0,2,0)}_{1} $ are $ 10 $ scalars $ tr( X^{A} \tilde{X}^{B}) $ and the complex conjugate $ tr( X_{A} \tilde{X}_{B})  $. The highest weight operator can be selected as $tr( X^{1} \tilde{X}^{1}) $ which is annihilated by $ 3 $ supercharges $\bar{\mathcal{Q}}^{12}  $, $\bar{\mathcal{Q}}^{13}   $ and $\bar{\mathcal{Q}}^{14}  $ based on (\ref{1033}) and (\ref{1033d}). The rest $ 3 $ supercharges generate the whole multiplet. The $ 6 $ dimension-$ 2 $ vectors $  [2]^{(1,0,0)}_{2}$ are extra R-symmetry currents $j^{AB}_{\mu }  $ in (\ref{jab}). The top component $[3]^{(0,0,0)}_{\frac{5}{2}} $ is the extra SUSY-current whose integration gives the supercharge $ \mathcal{Q} $ with the adjoint $   \mathcal{Q} ^{+}$.

When the extra SUSY-current multiplets exist, the $ \mathcal{N} =6  $ theory actually has $ 8 $ supercharges $\bar{\mathcal{Q}}^{AB}$, $  \bar{\mathcal{Q}}$ and $ \bar{\mathcal{Q}}^{+} $. According to \cite{tw15}, the 3d $  \mathcal{N} =8 $ SCFT has a $ 128+128 $ $ 1/2 $ BPS stress tensor multiplet with $ 15 +10+10$ chiral primary operators in $[0]^{(0,0,0,2)}_{1}$ representation. The whole multiplet is given by
\begin{equation}
[0]^{(0,0,0,2)}_{1}\stackrel{Q}{\longrightarrow}[1]^{(0,0,1,1) }_{\frac{3}{2}}\stackrel{Q}{\longrightarrow} [0]^{(0,0,2,0)}_{2}\oplus [2]^{(0,1,0,0)}_{2}\stackrel{Q}{\longrightarrow}[3]^{(1,0,0,0)}_{\frac{5}{2}}\stackrel{Q}{\longrightarrow}[4]^{(0,0,0,0)}_{3}\;.
\end{equation}
$[0]^{(0,0,0,2)}_{1} $ is composed of $ 15 $ scalars $ tr(X^{A}X_{B}-\frac{1}{4}\delta_{B}^{A}X^{C}X_{C}) $, $ 10 $ scalars $ tr( X^{A} \tilde{X}^{B}) $ and $ 10 $ scalars $ tr( X_{A} \tilde{X}_{B})  $, among which the highest weight operator can be selected as $tr( X^{1} \tilde{X}^{1})$. The action of the $ SO(8) $ R-charges gives the complete $[0]^{(0,0,0,2)}_{1}   $ representation. From (\ref{1033})-(\ref{1033de}), $tr(X^{1} \tilde{X}^{1})$ is a $ 1/2 $ BPS operator annihilated by $ 4 $ supercharges $\bar{\mathcal{Q}}^{12}  $, $\bar{\mathcal{Q}}^{13}   $, $\bar{\mathcal{Q}}^{14}  $ and $  \bar{\mathcal{Q}}  $. The action of the rest $ 4 $ supercharges gives the whole multiplet, among which, $ [2]^{(0,1,0,0)}_{2} $, $ [3]^{(1,0,0,0)}_{\frac{5}{2}} $ and $[4]^{(0,0,0,0)}_{3}  $ are $ SO(8) $ R-symmetry current, $ \mathcal{N} =8$ SUSY current and the stress tensor that is just $ [4]^{(0,0,0)}_{3}  $ in $ \mathcal{N} =6$ language.

The extra SUSY-current multiplets are built from $ \mathcal{M}_{\hat{a} \hat{b} }^{ab}(x)  $ and $ \mathcal{M}^{\hat{a} \hat{b} }_{ab}(x)  $, which is the only possibility when $ k=2 $. However, when $ k=1 $, it seems that $ \mathcal{M}'^{ab}_{\hat{a} \hat{b} }(x)  $ and $ \mathcal{M}'^{\hat{a} \hat{b} }_{ab}(x)  $ in (\ref{sasq}) will give another set of extra SUSY-current. The existence of two copies of $ \mathcal{N} =8$ supercurrents in $ k=1 $ ABJM model is discussed in \cite{tw14}.

Besides, $ 1/6 $ BPS higher-spin current multiplet with chiral primary operators in $[0]^{(1,0,0)}_{1}$ representation is also allowed in 3d $  \mathcal{N} =6 $ SCFT \cite{tw15}. The whole multiplet is given by 
\begin{equation}\label{1039i}
[0]^{(1,0,0)}_{1}\stackrel{Q}{\longrightarrow}[1]^{(0,0,0)\oplus (0,1,1)     }_{\frac{3}{2}}\stackrel{Q}{\longrightarrow} [0]^{(1,0,0)}_{2}\oplus [2]^{(0,0,2)\oplus (0,2,0)}_{2}\stackrel{Q}{\longrightarrow}[3]^{(0,1,1)}_{\frac{5}{2}}\stackrel{Q}{\longrightarrow}[4]^{(1,0,0)}_{3}\stackrel{Q}{\longrightarrow}[5]^{(0,0,0)}_{\frac{7}{2}}\;.
\end{equation}
The scalars $  \mathcal{M}^{ab}_{\hat{a}\hat{b}}  X_{a}^{A\hat{a}}X_{b}^{B\hat{b}}  $ are in the $ [0]^{(0,0,2)}_{1} $ representation. When $ k=1 $, using $\mathcal{M}''^{ab} _{\hat{a}\hat{b}} $ or $ \mathcal{M}'''^{ab}_{\hat{a}\hat{b}}  $ in (\ref{SAS}) and (\ref{SAS1}), $  \mathcal{M}''^{ab}_{\hat{a}\hat{b}}  X_{a}^{A\hat{a}}X_{b}^{B\hat{b}} $ or $  \mathcal{M}'''^{ab}_{\hat{a}\hat{b}}  X_{a}^{A\hat{a}}X_{b}^{B\hat{b}} $ in representation $ [0]^{(1,0,0)}_{1} $ can be obtained, which are annihilated by $ 1 $ supercharge. The rest $ 5 $ supercharges generate the complete multiplet. From $\mathcal{M}''^{ab} _{\hat{a}\hat{b}} ( \mathcal{M}'''^{ab}_{\hat{a}\hat{b}}   )$, one can only construct the separate bosonic currents
\begin{equation}
j^{AB}_{\mu \;X}=-\frac{ik}{4\pi}\mathcal{M}''^{ab}_{\hat{a} \hat{b} }[X_{a}^{A \hat{a}}(D_{\mu}X^{B})_{b}^{\hat{b}}-(D_{\mu}X^{A})_{a}^{\hat{a}}X_{b}^{B \hat{b}}]
\end{equation}
and fermionic currents 
\begin{equation}
j_{\mu \;AB\;\Psi }=\frac{k}{4\pi}\mathcal{M}''^{ab}_{\hat{a} \hat{b} }[ \bar{\Psi}_{Aa}^{\hat{a}}\gamma_{\mu} \Psi_{Bb}^{ \hat{b}}]\;,
\end{equation}
corresponding to $ [2]^{(0,0,2)}_{2} $ and $ [2]^{(0,2,0)}_{2} $ in (\ref{1039i}). Similarly to the Appendix \ref{AAA}, we may find that $ j^{AB}_{\mu \;X} $ and $j_{\mu \;AB\;\Psi }  $ are not conserved due to the fermionic interaction (\ref{712z}). This is expected, since otherwise, we will get the conserved higher-spin current which can only exist in free theory \cite{higher}. So the multiplet (\ref{1039i}) is unprotected and would not bring the additional symmetry.

Finally, 3d $  \mathcal{N} =6 $ SCFT can also have the free hypermultiplets exchanged by complex conjugation \cite{tw15}:
\begin{equation}
[0]^{(0,1,0)}_{\frac{1}{2}}\stackrel{Q}{\longrightarrow}[1]^{(0,0,1)}_{1} \;,\;\;\;\;\;\;
[0]^{(0,0,1)}_{\frac{1}{2}}\stackrel{Q}{\longrightarrow}[1]^{(0,1,0)}_{1}\;.
\end{equation}
In the $ k=1 $ ABJM model, $[0]^{(0,1,0)}_{\frac{1}{2}}  $ and $ [0]^{(0,0,1)}_{\frac{1}{2}} $, $  [1]^{(0,0,1)}_{1}$ and $ [1]^{(0,1,0)}_{1} $ are scalars $ x^{A} =\mathcal{M}^{a}_{\hat{a} }  X_{a}^{A\hat{a}} $ and $  x_{A} =\mathcal{M}_{a}^{\hat{a} }  X^{a}_{A\hat{a}}  $, spinors $   \psi_{A} =\mathcal{M}^{a}_{\hat{a} }  \Psi^{\hat{a}}_{Aa}   $ and $\psi^{A} =\mathcal{M}_{a}^{\hat{a} }  \Psi_{\hat{a}}^{Aa} $. From (\ref{911aa}), (\ref{911aaa}), (\ref{a3}) and (\ref{a4}), using (\ref{89i}), we have
\begin{equation}
 \delta x_{A}=i  \bar{\psi}^{B}  \eta_{AB} \;,\;\;\;\;\;\;\delta \psi_{A}=\gamma^{\mu} \eta_{AB}  \partial_{\mu}x^{B}\;,
\end{equation}
and 
\begin{equation}
\partial^{2} x_{A}=0 \;,\;\;\;\;\;\; \gamma^{\mu}\partial_{\mu}\psi_{A} =0\;.
\end{equation}
$( x , \psi)$ are indeed free. In $ U(N) $ SYM theories, the decoupled free sector is the trace of the fundamental fields, but here the monopole operators must be used.

\section{Conclusion and discussion}\label{cd}

The main result of the paper is composed by two parts. First, based on the original definition in \cite{Hoo}, we constructed the monopole operators and computed the contraction relations, classical conformal dimensions, supersymmetry transformations and the covariant derivatives. Second, with the concrete form of the monopole operators given, we studied their role in the global symmetry enhancement of the ABJM theory and proved several assumptions that were made on them to achieve the symmetry enhancement.

In Chern-Simons-matter theories with the nondynamical gauge fields, monopole operators commute with the supercharges, the covariant derivative operators, and some particular field-dependent gauge variation operators. The ordinary gauge transformations also commute with these gauge invariant operators, while the monopole operators are just the singular gauge transformations whose singularity does not have the manifestation here. As a result, except for changing the matter field's gauge representation, monopole operators do not have the side effect when combined with the matter, which makes them the suitable ingredients in the symmetry enhancement.

5d gauge theories also contain a conserved topological current $ J^{\mu}=\frac{1}{32\pi^{2}}\epsilon^{\mu\nu\lambda\alpha\beta}tr(F_{\nu\lambda}F_{\alpha\beta}) $. The conserved charge 
\begin{equation}\label{1qaz}
Q_{I}=\frac{1}{32\pi^{2}}\int d^{4}y\; \epsilon^{ijkl}tr(F_{ij}F_{kl})
\end{equation}
is the instanton number, and the corresponding global symmetry is the $ U(1)_{I} $ symmetry. As the higher-dimensional analogues of the 3d monopole operators, instanton operators are local disorder operators creating instanton number on a $ S^{4} $ surrounding their insertion point and could be defined by specifying the field configurations carrying the nonvanishing instanton number on $ S^{4} $ \cite{1a,2a,2ac}. Global symmetry enhancement may occur at the UV fixed point of the 5d gauge theories \cite{1b}. The original global symmetry algebra $ \mathcal{R} \oplus Q_{I}$ of the theory can be enhanced to $ \hat{\mathcal{R}} = \mathcal{R}\oplus Q_{I} \oplus  \mathcal{R}_{\mathrm{off-diag}}  $ if the currents for $ \mathcal{R}_{\mathrm{off-diag}}  $ can be built from the instanton operators and are conserved \cite{1c, 2c, 3c, 4c}. In parallel with the discussion for monopole operators, one may write down the instanton operators, construct the $ \mathcal{R}_{\mathrm{off-diag}}  $ currents, and investigate their conservation.

\section*{Acknowledgments}

The work is supported in part by NSFC under Grant No. 11605049.


\begin{appendix}

\section{Solitons and soliton operators}\label{AAA1}

In a $ D$-dimensional gauge theory with the gauge group $ G $, the finite-action gauge field configuration should satisfy 
\begin{equation}
A_{i} \rightarrow i u^{-1} \partial_{i} u\;,\;\;\;\;\;\;\;u \in G
\end{equation}
at the spatial boundary $\partial R^{D-1} \cong S_{\infty}^{D-2}$, $ i=1,\cdots,D-1 $. So, the finite-action configurations provide maps from $ S_{\infty}^{D-2} $ to $ G $, which is labeled by the homotopy group $  \Pi_{D-2}(G)   $. Two maps are in the same homotopy class if they can be continuously deformed into each other. The surface integral
\begin{equation}
\int_{S_{\infty}^{D-2}} \; tr[(u^{-1} du)^{D-2}]
\end{equation}
counts how many times the group wraps itself around $  S_{\infty}^{D-2} $ and is simply $ 0 $ when $  \Pi_{D-2}(G) =0 $.

When $ D=3 $, $ G=U(N) $, 2d gauge fields are classified by the homotopy group $ \Pi_{1}(U(N) ) \cong \mathbb{Z} $. The integer $ q \in  \mathbb{Z}$ is just the vortex charge (\ref{1q})
\begin{equation}
Q =\frac{1}{4\pi} \int  d^{2}x \; \epsilon^{ij}tr F_{ij}  =\frac{1}{2\pi} \int_{S_{\infty}^{1}}  dS_{i} \; \epsilon^{ij} tr A_{j} =\frac{i}{2\pi} \int_{S_{\infty}^{1}}  dS_{i} \; \epsilon^{ij} tr (u^{-1} \partial_{j} u)\;.
\end{equation}
If $ \vert A_{i}  \rangle   $ is in a homotopy class labeled by $ q $, then
\begin{equation}
Q \vert A_{i}  \rangle = q \vert A_{i}  \rangle\;. 
\end{equation}
Vortex operators are ``large'' gauge transformations that could make the states in different homotopy classes transform into each other. Consider a gauge transformation $ U $ with $ UA_{i}U^{-1}= g^{-1}A_{i}g+ig^{-1}\partial_{i}g $; if 
\begin{equation}
\frac{i}{2\pi} \int_{S_{\infty}^{1}}  dS_{i} \; \epsilon^{ij} tr (g^{-1} \partial_{j} g)=k\;,
\end{equation}
then 
\begin{equation}
UQ  U^{-1}=Q+\frac{i}{2\pi} \int_{S_{\infty}^{1}}  dS_{i} \; \epsilon^{ij} tr (g^{-1} \partial_{j} g)=Q+k\;.
\end{equation}
$[Q,U]=-kU  $. $ U  $ is a vortex operator (monopole operator) carrying $ -k $ vortex charges.  
\begin{equation}\label{a77}
Q  U\vert A_{i}  \rangle = (q-k)  U\vert A_{i}  \rangle\;.
\end{equation}
$  U\vert A_{i}  \rangle $ is in a homotopy class labeled by $ q -k$. The ordinary gauge transformation should have $ k=0 $. When $ k\neq 0 $, $ g $ must be singular in at least one point, which is the location of the vortex operator. Such $ g $ is gauge equivalent to $ \Omega_{\vec{m}} (\omega)$ in section \ref{mr}.

When $ D=5 $, $G=U(N)  $ or $ SU(N) $ and $ N\geq 2 $, 4d gauge fields are classified by the homotopy group $  \Pi_{3}(G) \cong \mathbb{Z} $. The integer $ q \in  \mathbb{Z}$ is the instanton number (\ref{1qaz})
\begin{eqnarray}
  Q_{I}   &=&\frac{1}{32\pi^{2}}\int d^{4}y\; \epsilon^{ijkl}tr(F_{ij}F_{kl})=\frac{1}{8\pi^{2}}\epsilon^{ijkl}\int_{S_{\infty}^{3}} d^{3}S_{i}\;tr[A_{j}(\partial_{k}A_{l}-\frac{2i}{3} A_{k}A_{l})]\\ &=& \frac{1}{24\pi^{2}}\epsilon^{ijkl}\int_{S_{\infty}^{3}} d^{3}S_{i}\;
 (u^{-1}\partial_{j}u)(u^{-1}\partial_{k}u)(u^{-1}\partial_{l}u) \;.
\end{eqnarray}
For a gauge transformation $ U $ with $ UA_{i}U^{-1}= g^{-1}A_{i}g+ig^{-1}\partial_{i}g $, if 
\begin{equation}
 \frac{1}{24\pi^{2}}\epsilon^{ijkl}\int_{S_{\infty}^{3}} d^{3}S_{i}\;
 (g^{-1}\partial_{j}g)(g^{-1}\partial_{k}g)(g^{-1}\partial_{l}g) =k\;,
\end{equation}
then 
\begin{equation}
UQ_{I}U^{-1}=Q_{I}+ \frac{1}{24\pi^{2}}\epsilon^{ijkl}\int_{S_{\infty}^{3}} d^{3}S_{i}\;
 (g^{-1}\partial_{j}g)(g^{-1}\partial_{k}g)(g^{-1}\partial_{l}g) =Q_{I}+k\;.
\end{equation}
$[Q_{I},U]=-kU  $. $ U  $ is an instanton operator with the instanton number $ -k $. For $ \vert A_{i}  \rangle $ in a homotopy class labeled by $ q $, $  U \vert A_{i}  \rangle $ is in a homotopy class labeled by $ q -k$. The ordinary gauge transformations have $ k=0 $. When $ k\neq 0 $, the action of $ U $ must be singular in at least one point that is the location of the instanton operator.

\section{Current conservation equation of $j^{AB}_{\mu }  $}\label{AAA}

The enhancement of the global symmetry from $ SU(4) \times U(1)_{J}$ to $ SO(8) $ requires the conservation of the $ 12 $ currents 
\begin{equation}
j^{AB}_{\mu }=-\frac{ik}{4\pi}\mathcal{M}_{\hat{a} \hat{b} }^{ab}[X_{a}^{A \hat{a}}(D_{\mu}X^{B})_{b}^{\hat{b}}-(D_{\mu}X^{A})_{a}^{\hat{a}}X_{b}^{B \hat{b}}+\frac{i}{2}\epsilon^{ABCD} \bar{\Psi}_{Ca}^{\hat{a}}\gamma_{\mu} \Psi_{Db}^{ \hat{b}}]
\end{equation}
and the adjoints $ j_{\mu AB} $.

Direct calculation gives
\begin{eqnarray}\label{A2}
\nonumber \partial^{\mu}j^{AB}_{\mu }&=&-\frac{ik}{4\pi}(D^{\mu}\mathcal{M})_{\hat{a} \hat{b} }^{ab}[X_{a}^{A \hat{a}}(D_{\mu}X^{B})_{b}^{\hat{b}}-(D_{\mu}X^{A})_{a}^{\hat{a}}X_{b}^{B \hat{b}}+\frac{i}{2}\epsilon^{ABCD} \bar{\Psi}_{Ca}^{\hat{a}}\gamma_{\mu} \Psi_{Db}^{ \hat{b}}]  \\ &-&\frac{ik}{4\pi}\mathcal{M}_{\hat{a} \hat{b} }^{ab}[X_{a}^{A \hat{a}}(D^{2}X^{B})_{b}^{\hat{b}}-(D^{2}X^{A})_{a}^{\hat{a}}X_{b}^{B \hat{b}}+i\epsilon^{ABCD} \bar{\Psi}_{Ca}^{\hat{a}}\gamma_{\mu}(D^{\mu} \Psi_{D})_{b}^{ \hat{b}}]  \;.
\end{eqnarray}
From the action (\ref{69i})-(\ref{613i}), we have
\begin{equation}\label{a3}
\frac{k}{2\pi}D^{2}X^{B}=R^{B}_{1}+R^{B}_{2}+R^{B}_{3}
\end{equation}
with 
\begin{eqnarray}
 \nonumber R^{B}_{1} &=&2 i\epsilon^{BCDE}\bar{\Psi}_{C}X_{D}\Psi_{E} \;,\\ \nonumber R^{B}_{2} &=&  i(\bar{\Psi}_{C} \Psi^{C}X^{B}  -  X^{B} \bar{\Psi}^{C} \Psi_{C}+2X^{C} \bar{\Psi}^{B} \Psi_{C}-2 \bar{\Psi}_{C} \Psi^{B}X^{C}   ) \;,\\
\nonumber R^{B}_{3} &=&2( X^{B}X_{C}X^{D}X_{D}X^{C}+X^{C}X_{D}X^{D}X_{C}X^{B}  -2X^{C}X_{D}X^{B}X_{C}X^{D}    )   \\ \nonumber&-& (X^{B}X_{C}X^{C}X_{D}X^{D}+X^{C}X_{C}X^{D}X_{D}X^{B} -2X^{C}X_{C}X^{B}X_{D}X^{D})
\;,
\end{eqnarray}
and
\begin{equation}\label{a4}
\frac{k}{2\pi} \epsilon^{ABCD}   \gamma^{\mu}D_{\mu}\Psi_{D} =R^{ABC}_{1}+R^{ABC}_{2}
\end{equation}
with 
\begin{eqnarray}
\nonumber R^{ABC}_{1}  &=& \epsilon^{ABCD}( X^{E} X_{E}  \Psi_{D} -   \Psi_{D} X_{E}X^{E} -2X^{E} X_{D} \Psi_{E}+2 \Psi_{E} X_{D}X^{E} )
\\  \nonumber R^{ABC}_{2}&=& 2 (X^{A}\Psi^{C}X^{B}-X^{A}\Psi^{B}X^{C}  +X^{B}\Psi^{A}X^{C}-X^{B}\Psi^{C}X^{A}+ X^{C}\Psi^{B}X^{A}  - X^{C}\Psi^{A}X^{B}  ) \;. 
\end{eqnarray}
With (\ref{a3}) and (\ref{a4}) plugged in, the second line of (\ref{A2}) becomes 
\begin{equation}
-\frac{i}{2}[\mathcal{M}(X^{[A} R^{B]}_{1}+X^{[A} R^{B]}_{2}+X^{[A} R^{B]}_{3})+i
\mathcal{M}\bar{\Psi}_{C}(R^{ABC}_{1}+R^{ABC}_{2})]\;.
\end{equation}

We may introduce the gauge variation $ W^{C}_{D} $, the action of which on $ X_{a}^{ B\hat{a}} $ is  
\begin{equation}
(W^{C}_{D} X^{B})_{a}^{ \hat{a}}\equiv (X^{C}X_{D})^{\hat{a}}_{\hat{b}}X_{a}^{B\hat{b}}-X^{B\hat{a}}_{b}(X_{D}X^{C})_{a}^{b}\;.
\end{equation}
$R^{B}_{3}  $ could be written as 
\begin{equation}
R^{B}_{3} = 2(W^{D}_{C} (W^{C}_{D} X^{B}))-(W^{C}_{C}(W^{D}_{D} X^{B}))\;.
\end{equation}
We have
\begin{eqnarray}
\nonumber \mathcal{M} X^{A}R^{B}_{3}&=&2\mathcal{M} X^{A} (W^{D}_{C} (W^{C}_{D} X^{B}))-\mathcal{M} X^{A} (W^{C}_{C}(W^{D}_{D} X^{B}))\\\nonumber &=&-2(W^{D}_{C} \mathcal{M} )X^{A} (W^{C}_{D} X^{B})-2\mathcal{M} (W^{D}_{C} X^{A})  (W^{C}_{D} X^{B})   \\ &+& (W^{C}_{C} \mathcal{M} )X^{A} (W^{D}_{D} X^{B})+\mathcal{M} (W^{C}_{C} X^{A}) (W^{D}_{D} X^{B})\;,
\end{eqnarray}
and 
\begin{eqnarray}\label{a10}
\nonumber \mathcal{M} X^{[A}R^{B]}_{3} &=&-2(W^{D}_{C} \mathcal{M} )X^{A} (W^{C}_{D} X^{B})+(W^{C}_{C} \mathcal{M} )X^{A} (W^{D}_{D} X^{B})\\ &+&2(W^{D}_{C} \mathcal{M} )X^{B} (W^{C}_{D} X^{A})-(W^{C}_{C} \mathcal{M} )X^{B} (W^{D}_{D} X^{A})\;.
\end{eqnarray}

Next, consider $i\mathcal{M}\bar{\Psi}_{C}R^{ABC}_{1}  $, whose explicit form is
\begin{equation}
i\mathcal{M}\bar{\Psi}_{C}R^{ABC}_{1}=i
 \epsilon^{ABCD}\mathcal{M}\bar{\Psi}_{C}( X^{E} X_{E}  \Psi_{D} -   \Psi_{D} X_{E}X^{E} -2X^{E} X_{D} \Psi_{E}+2 \Psi_{E} X_{D}X^{E} )\;.
\end{equation}
Construct a gauge variation $ V_{CD} $, the actions of which on $ X_{a}^{B \hat{a}} $ and $ \bar{\Psi}_{B} $ are given by
\begin{equation}
(V_{CD} X^{B})_{a}^{ \hat{a}} \equiv (\Psi_{C} X_{D})^{\hat{a}}_{\hat{b}}X_{a}^{B\hat{b}}-X^{B\hat{a}}_{b}(X_{D} \Psi_{C})_{a}^{b}
\end{equation}
and 
\begin{equation}
(V_{CD} \bar{\Psi}_{B})_{a}^{ \hat{a}} \equiv (\bar{\Psi}_{C} X_{D})^{\hat{a}}_{\hat{b}}\Psi_{Ba}^{\hat{b}}-\bar{\Psi}^{\hat{a}}_{B b}(X_{D} \Psi_{C})_{a}^{b}\;,
\end{equation}
respectively. $ i\mathcal{M}\bar{\Psi}_{C}R^{ABC}_{1}   $ could be written as 
\begin{eqnarray}
\nonumber i\mathcal{M}\bar{\Psi}_{C}R^{ABC}_{1} &=&i \epsilon^{ABCD}\mathcal{M}\bar{\Psi}_{C}[ 2 (V_{ED} X^{E})-(V_{DE} X^{E}) ] \\ \nonumber &=& i\epsilon^{ABCD}X^{E}[\bar{\Psi}_{C} (V_{DE} \mathcal{M})-2\bar{\Psi}_{C}( V_{ED} \mathcal{M})+\mathcal{M}(V_{DE} \bar{\Psi}_{C}-2 V_{ED}\bar{\Psi}_{C})  ]  \;,
\end{eqnarray}
where
\begin{equation}
V_{DE} \bar{\Psi}_{C}-2 V_{ED}\bar{\Psi}_{C}=\bar{\Psi}_{D} X_{E} \Psi_{C}- \bar{\Psi}_{C}X_{E} \Psi_{D} -2 \bar{\Psi}_{E} X_{D} \Psi_{C}+2 \bar{\Psi}_{C}X_{D} \Psi_{E}\;.
\end{equation}
On the other hand, 
\begin{eqnarray}
\nonumber \mathcal{M}X^{[A} R^{B]}_{1} &=&2 i \mathcal{M}X^{[A}  \epsilon^{B]CDE}\bar{\Psi}_{C}X_{D}\Psi_{E}  \\ &=&i \mathcal{M}\epsilon^{ABCD} X^{E} (2 \bar{\Psi}_{E}X_{D}\Psi_{C}  - 2\bar{\Psi}_{C}X_{D}\Psi_{E} +\bar{\Psi}_{C}X_{E}\Psi_{D}-\bar{\Psi}_{D}X_{E}\Psi_{C} ) \;.
\end{eqnarray}
So
\begin{equation}\label{a16}
\mathcal{M}X^{[A} R^{B]}_{1} +i\mathcal{M}\bar{\Psi}_{C}R^{ABC}_{1} = i\epsilon^{ABCD}X^{E}[ \bar{\Psi}_{C}(V_{DE} \mathcal{M})-2\bar{\Psi}_{C}( V_{ED} \mathcal{M})]\;.
\end{equation}

Finally, consider $ \mathcal{M}X^{[A} R^{B]}_{2}  $. The actions of the gauge variation $ U^{CB} $ on $ \Psi_{Aa}^{ \hat{a}} $ and $ X_{a}^{A \hat{a}}   $ are defined as 
\begin{equation}
(U^{CB}  \Psi_{A})_{a}^{ \hat{a}} \equiv (X^{C} \bar{\Psi}^{B})^{\hat{a}}_{\hat{b}}\Psi_{A a}^{\hat{b}}-\bar{\Psi}^{\hat{a}}_{A b}(\Psi^{B}X^{C})_{a}^{b}
\end{equation}
and 
\begin{equation}
(U^{CB}  X^{A})_{a}^{ \hat{a}} \equiv (X^{C} \bar{\Psi}^{B})^{\hat{a}}_{\hat{b}}X_{ a}^{A \hat{b}}-X^{A\hat{a}}_{ b}(\bar{\Psi}^{B}X^{C})_{a}^{b}\;,
\end{equation}
respectively. Then, 
\begin{equation}
R^{B}_{2} =  i[2(U^{CB}  \Psi_{C}) - (U^{BC}  \Psi_{C}) ]\;.
\end{equation}
$\mathcal{M}X^{[A} R^{B]}_{2}  $ becomes
\begin{eqnarray}
\nonumber  \mathcal{M}X^{[A} R^{B]}_{2}&=&i \mathcal{M}\{X^{A} [2(U^{CB}  \Psi_{C}) - (U^{BC}  \Psi_{C}) ]-X^{B} [2(U^{CA}  \Psi_{C}) - (U^{AC}  \Psi_{C}) ]\}\\\nonumber &=&i[(U^{BC}  \mathcal{M})  X^{A}  -2(U^{CB}   \mathcal{M} )X^{A}   +2(U^{CA}  \mathcal{M} )X^{B} -( U^{AC}  \mathcal{M} )X^{B} ]\Psi_{C}   \\ &+& i\mathcal{M} [(U^{BC}   X^{A} )  - ( U^{AC} X^{B} ) 
+2 (U^{CA}  X^{B})  -2( U^{CB}  X^{A} ) ]\Psi_{C}  \;,
\end{eqnarray}
where 
\begin{eqnarray}
\nonumber && (U^{BC}   X^{A} )  - ( U^{AC} X^{B} ) 
+2 (U^{CA}  X^{B})  -2( U^{CB}  X^{A} ) \\ \nonumber &=&2 (X^{B} \bar{\Psi}^{C}X^{A}-X^{A} \bar{\Psi}^{C}X^{B}
+ X^{C} \bar{\Psi}^{A}X^{B}-X^{B} \bar{\Psi}^{A}X^{C}-X^{C} \bar{\Psi}^{B}X^{A}+X^{A} \bar{\Psi}^{B}X^{C})
\;.\\
\end{eqnarray}
On the other hand, 
\begin{eqnarray}
\nonumber &&i
\mathcal{M}\bar{\Psi}_{C}R^{ABC}_{2} \\ \nonumber &=&2i
\mathcal{M}\bar{\Psi}_{C} (X^{A}\Psi^{C}X^{B}-X^{A}\Psi^{B}X^{C}  +X^{B}\Psi^{A}X^{C}-X^{B}\Psi^{C}X^{A}+ X^{C}\Psi^{B}X^{A}  - X^{C}\Psi^{A}X^{B}  ) \\ \nonumber &=& 2i
\mathcal{M} (X^{A}\bar{\Psi}^{C}X^{B}-X^{A}\bar{\Psi}^{B}X^{C}  +X^{B}\bar{\Psi}^{A}X^{C}-X^{B}\bar{\Psi}^{C}X^{A}+ X^{C}\bar{\Psi}^{B}X^{A}  - X^{C}\bar{\Psi}^{A}X^{B}  )\Psi_{C}
\;.\\
\end{eqnarray}
We get
\begin{equation}\label{a23}
 \mathcal{M}X^{[A} R^{B]}_{2}+     i
\mathcal{M}\bar{\Psi}_{C}R^{ABC}_{2} =i[(U^{BC}  \mathcal{M})  X^{A}  -2(U^{CB}   \mathcal{M} )X^{A}   +2(U^{CA}  \mathcal{M} )X^{B} -( U^{AC}  \mathcal{M} )X^{B} ]\Psi_{C} 
\;.
\end{equation}

With (\ref{a10}), (\ref{a16}) and (\ref{a23}) combined together, 
\begin{eqnarray}
\nonumber &&\mathcal{M}(X^{[A} R^{B]}_{1}+X^{[A} R^{B]}_{2}+X^{[A} R^{B]}_{3})+i
\mathcal{M}\bar{\Psi}_{C}(R^{ABC}_{1}+R^{ABC}_{2}) \\ \nonumber &=&2(W^{D}_{C} \mathcal{M} )X^{B} (W^{C}_{D} X^{A})-2(W^{D}_{C} \mathcal{M} )X^{A} (W^{C}_{D} X^{B})+(W^{C}_{C} \mathcal{M} )X^{A} (W^{D}_{D} X^{B})\\ \nonumber &-&(W^{C}_{C} \mathcal{M} )X^{B} (W^{D}_{D} X^{A}) + i\epsilon^{ABCD}X^{E} \bar{\Psi}_{C}(V_{DE} \mathcal{M})-2 i\epsilon^{ABCD}X^{E}\bar{\Psi}_{C}
( V_{ED} \mathcal{M})\\  \nonumber &+& 
i(U^{BC}  \mathcal{M})  X^{A}\Psi_{C}   -2i(U^{CB}   \mathcal{M} )X^{A}  \Psi_{C}  +2i(U^{CA}  \mathcal{M} )X^{B}\Psi_{C}  -i( U^{AC}  \mathcal{M} )X^{B} \Psi_{C} 
\;.\\
\end{eqnarray}
The conservation condition for $j^{AB}_{\mu }  $ is 
\begin{eqnarray}\label{app}
\nonumber \partial^{\mu}j^{AB}_{\mu }&=&-\frac{ik}{4\pi}(D^{\mu}\mathcal{M})[X^{A }(D_{\mu}X^{B})-(D_{\mu}X^{A})X^{B}+\frac{i}{2}\epsilon^{ABCD} \bar{\Psi}_{C}\gamma_{\mu} \Psi_{D}] \\ \nonumber &-&(W^{D}_{C} \mathcal{M} )[X^{B} (W^{C}_{D} X^{A})-X^{A} (W^{C}_{D} X^{B})]-\frac{1}{2}(W^{C}_{C} \mathcal{M} )[X^{A} (W^{D}_{D} X^{B})-X^{B} (W^{D}_{D} X^{A})]\\ \nonumber &-&  \frac{i}{2} \bar{\Psi}_{C}(V_{DE} \mathcal{M})(\epsilon^{ABCD}X^{E}-2 \epsilon^{ABCE}X^{D})
\\ \nonumber  &-& 
\frac{i}{2}[(U^{BC}  \mathcal{M})  X^{A}  -( U^{AC}  \mathcal{M} )X^{B}  +2(U^{CA}  \mathcal{M} )X^{B}    -2(U^{CB}   \mathcal{M} )X^{A} ] \Psi_{C} 
\\  &=&0\;.
\end{eqnarray}

\end{appendix}

\end{document}